\documentclass[%
nofootinbib,
 amsmath,amssymb,
 aps,
 prd,
11pt,a4paper]{revtex4-2}
\usepackage[british]{babel}
\usepackage[lmargin=2.25cm,rmargin=2.25cm,tmargin=2.5cm,bmargin=3.6cm]{geometry}

\usepackage{graphicx}% Include figure files
\usepackage{dcolumn}% Align table columns on decimal point
\usepackage{bm}

\usepackage{amsfonts}
\usepackage{amsthm}
\usepackage[dvipsnames]{xcolor}
\usepackage{physics}
\usepackage{url}
\usepackage{tensor}
\usepackage{float}
\usepackage{mathrsfs}
\usepackage{mathtools}
\usepackage{bm}
\usepackage{cancel}
\usepackage{tikz}
\usepackage[shortlabels]{enumitem}
\usepackage{mdframed}
\usepackage[pdftex,hidelinks,colorlinks=true, allcolors=blue]{hyperref}
\usepackage{tikz-cd}

\usepackage{dcolumn}

\numberwithin{equation}{section}

\DeclareSymbolFont{matha}{OML}{txmi}{m}{it}% txfonts
\DeclareMathSymbol{v}{\mathord}{matha}{118}

\def\fheq{\overset{\mathcal H^+}{=}}

\begin{document}

\title{Gravitational focusing and horizon entropy for higher-spin fields}

\author{Zihan~Yan}
\email{zy286@cam.ac.uk}
\affiliation{DAMTP, Centre for Mathematical Sciences, University of Cambridge\\Wilberforce Road, Cambridge, U.K. CB3 0WA}

\date{\today}

\begin{abstract}
    Previously, the Raychaudhuri equation and the focusing theorem in General Relativity were generalised to diffeomorphism-invariant theories of gravity coupled to scalar and vector fields on linearly perturbed Killing horizons. The Wall entropy can be extracted from the generalised focusing equation and it satisfies the first and the second laws of thermodynamics. In this paper, we further extend the discussion of gravitational focusing on the horizon to include arbitrary bosonic fields with spin $s \geq 2$. These higher-spin fields introduce indefinite terms into the generalised focusing equation, obstructing the proof of the focusing theorem and the existence of an increasing horizon entropy. To resolve this issue, we propose a higher-spin focusing condition that eliminates these indefinite terms, thereby restoring the focusing theorem and the associated thermodynamic laws. We speculate that the focusing condition could be a necessary condition for the physical consistency of higher-spin theories.
\end{abstract}

\maketitle\vspace{-0.4cm}
\tableofcontents
    
\section{Introduction}

\subsection{Focusing theorem and Bekenstein-Hawking entropy in GR}
In General Relativity (GR), light rays exhibit a riveting property: neighbouring light beams always tend to focus with each other, given that the matter sector satisfies the null energy condition. Qualitatively, two adjacent light rays can start anti-focused, yet they may bend to converge with each other and become parallel at a late time. If they ever start to meet, then this process is irreversible, and a conjugate point where they intersect is inevitable at a finite affine parameter. The focusing property of light rays is a manifestation that gravity is an attractive force, and it forebodes the inevitability of singularities when there exist trapped surfaces in non-compact space, leading to the famous Hawking-Penrose singularity theorem \cite{Penrose:1964wq,Hawking:1973uf}. 

\begin{figure}[H]
    \centering

    \tikzset{every picture/.style={line width=0.75pt}} %set default line width to 0.75pt        

    \begin{tikzpicture}[x=0.75pt,y=0.75pt,yscale=-1,xscale=1]

    %Curve Lines [id:da32879594510349475] 
    \draw    (110,70) .. controls (110,107.14) and (120.6,152.43) .. (130,170) ;
    \draw [shift={(114.14,115.05)}, rotate = 79.83] [fill={rgb, 255:red, 0; green, 0; blue, 0 }  ][line width=0.08]  [draw opacity=0] (8.93,-4.29) -- (0,0) -- (8.93,4.29) -- cycle    ;
    %Curve Lines [id:da6286899754451338] 
    \draw    (160,70) .. controls (160,107.14) and (149.8,152.43) .. (140,170) ;
    \draw [shift={(155.97,115.05)}, rotate = 99.98] [fill={rgb, 255:red, 0; green, 0; blue, 0 }  ][line width=0.08]  [draw opacity=0] (8.93,-4.29) -- (0,0) -- (8.93,4.29) -- cycle    ;
    %Curve Lines [id:da5714876361138992] 
    \draw    (207.38,170) .. controls (209.18,143.86) and (209.58,113.35) .. (227.38,88.35) ;
    \draw [shift={(212.67,122.45)}, rotate = 101.67] [fill={rgb, 255:red, 0; green, 0; blue, 0 }  ][line width=0.08]  [draw opacity=0] (8.93,-4.29) -- (0,0) -- (8.93,4.29) -- cycle    ;
    %Curve Lines [id:da6383332407877707] 
    \draw    (247.38,170) .. controls (245.58,143.86) and (243.58,112.78) .. (227.38,88.35) ;
    \draw [shift={(241.51,122.63)}, rotate = 77.9] [fill={rgb, 255:red, 0; green, 0; blue, 0 }  ][line width=0.08]  [draw opacity=0] (8.93,-4.29) -- (0,0) -- (8.93,4.29) -- cycle    ;
    %Straight Lines [id:da9887027571698694] 
    \draw    (227.38,88.35) ;
    \draw [shift={(227.38,88.35)}, rotate = 0] [color={rgb, 255:red, 0; green, 0; blue, 0 }  ][fill={rgb, 255:red, 0; green, 0; blue, 0 }  ][line width=0.75]      (0, 0) circle [x radius= 2.01, y radius= 2.01]   ;
    %Shape: Rectangle [id:dp719037674186104] 
    \draw  [draw opacity=0] (40,70) -- (110,70) -- (110,110) -- (40,110) -- cycle ;
    %Curve Lines [id:da06818565972554924] 
    \draw    (232.63,71.75) .. controls (232.13,77) and (230.5,83) .. (227.38,88.35) ;
    %Curve Lines [id:da5856579415330092] 
    \draw    (222,72) .. controls (222.25,75.75) and (223,83.5) .. (227.38,88.35) ;

    % Text Node
    \draw (285.25,87) node   [align=left] {conjugate point};

    \end{tikzpicture}

    \caption{Left: defocused light rays paralleling. Right: focused light rays intersecting.}
\end{figure}

Quantitatively, one can describe the evolution of a null geodesic congruence (a family of geodesics), especially the focusing property, using the \emph{Raychaudhuri equation} (originally in \cite{PhysRev.98.1123})
\begin{equation}
    \partial_v \theta = - \frac{1}{D-2} \theta^2 - \sigma_{ab} \sigma^{ab} - R_{vv}
\end{equation}
where $D$ is the spacetime dimension, $v$ is an affine null parameter and $k = \partial_v$ is future-directed, $\theta$ is the expansion of the congruence, $\sigma_{ab}$ is the shear, and $R_{vv}$ is the null-null component of Ricci tensor. Particularly, the expansion can be expressed as 
\begin{equation}
    \theta = \frac{1}{\mathscr{A}} \partial_v \mathscr{A}
\end{equation}
where $\mathscr{A}$ is the infinitesimal codimension-2 area element spanned by (a fixed number of) nearby geodesics, which gives a direct measure of how close to each other the geodesics are. 

\begin{figure}[H]
    \centering

\tikzset{every picture/.style={line width=0.75pt}} %set default line width to 0.75pt        

\begin{tikzpicture}[x=0.75pt,y=0.75pt,yscale=-1,xscale=1]

%Curve Lines [id:da2879188742787897] 
\draw    (65,50.5) .. controls (69.5,98) and (73,119) .. (103.5,152.5) ;
\draw [shift={(103.5,152.5)}, rotate = 47.68] [color={rgb, 255:red, 0; green, 0; blue, 0 }  ][fill={rgb, 255:red, 0; green, 0; blue, 0 }  ][line width=0.75]      (0, 0) circle [x radius= 1.34, y radius= 1.34]   ;
\draw [shift={(73.3,103.02)}, rotate = 71.65] [fill={rgb, 255:red, 0; green, 0; blue, 0 }  ][line width=0.08]  [draw opacity=0] (3.57,-1.72) -- (0,0) -- (3.57,1.72) -- cycle    ;
\draw [shift={(65,50.5)}, rotate = 84.59] [color={rgb, 255:red, 0; green, 0; blue, 0 }  ][fill={rgb, 255:red, 0; green, 0; blue, 0 }  ][line width=0.75]      (0, 0) circle [x radius= 1.34, y radius= 1.34]   ;
%Curve Lines [id:da443771732093057] 
\draw    (91,68.5) .. controls (93,111.5) and (98.5,139) .. (114,160.5) ;
\draw [shift={(114,160.5)}, rotate = 54.21] [color={rgb, 255:red, 0; green, 0; blue, 0 }  ][fill={rgb, 255:red, 0; green, 0; blue, 0 }  ][line width=0.75]      (0, 0) circle [x radius= 1.34, y radius= 1.34]   ;
\draw [shift={(95.58,113.75)}, rotate = 78.67] [fill={rgb, 255:red, 0; green, 0; blue, 0 }  ][line width=0.08]  [draw opacity=0] (3.57,-1.72) -- (0,0) -- (3.57,1.72) -- cycle    ;
\draw [shift={(91,68.5)}, rotate = 87.34] [color={rgb, 255:red, 0; green, 0; blue, 0 }  ][fill={rgb, 255:red, 0; green, 0; blue, 0 }  ][line width=0.75]      (0, 0) circle [x radius= 1.34, y radius= 1.34]   ;
%Curve Lines [id:da6796071384604101] 
\draw    (111,39) .. controls (118.5,86.5) and (118.5,110.5) .. (117.5,147) ;
\draw [shift={(117.5,147)}, rotate = 91.57] [color={rgb, 255:red, 0; green, 0; blue, 0 }  ][fill={rgb, 255:red, 0; green, 0; blue, 0 }  ][line width=0.75]      (0, 0) circle [x radius= 1.34, y radius= 1.34]   ;
\draw [shift={(117.03,90.1)}, rotate = 86.62] [fill={rgb, 255:red, 0; green, 0; blue, 0 }  ][line width=0.08]  [draw opacity=0] (3.57,-1.72) -- (0,0) -- (3.57,1.72) -- cycle    ;
\draw [shift={(111,39)}, rotate = 81.03] [color={rgb, 255:red, 0; green, 0; blue, 0 }  ][fill={rgb, 255:red, 0; green, 0; blue, 0 }  ][line width=0.75]      (0, 0) circle [x radius= 1.34, y radius= 1.34]   ;
%Curve Lines [id:da21529394139719793] 
\draw    (134,50.5) .. controls (139.5,86.5) and (143,117.5) .. (130,156) ;
\draw [shift={(130,156)}, rotate = 108.66] [color={rgb, 255:red, 0; green, 0; blue, 0 }  ][fill={rgb, 255:red, 0; green, 0; blue, 0 }  ][line width=0.75]      (0, 0) circle [x radius= 1.34, y radius= 1.34]   ;
\draw [shift={(138.94,100.78)}, rotate = 90.69] [fill={rgb, 255:red, 0; green, 0; blue, 0 }  ][line width=0.08]  [draw opacity=0] (3.57,-1.72) -- (0,0) -- (3.57,1.72) -- cycle    ;
\draw [shift={(134,50.5)}, rotate = 81.31] [color={rgb, 255:red, 0; green, 0; blue, 0 }  ][fill={rgb, 255:red, 0; green, 0; blue, 0 }  ][line width=0.75]      (0, 0) circle [x radius= 1.34, y radius= 1.34]   ;
%Shape: Parallelogram [id:dp14117427805324034] 
\draw   (94.7,135) -- (157,135) -- (130.3,175) -- (68,175) -- cycle ;
%Shape: Parallelogram [id:dp9894428643556121] 
\draw   (71,29.5) -- (171,29.5) -- (129,78.5) -- (29,78.5) -- cycle ;
%Shape: Polygon Curved [id:ds5298602116795041] 
\draw  [draw opacity=0][fill={rgb, 255:red, 74; green, 144; blue, 226 }  ,fill opacity=0.25 ] (111,39) .. controls (114.5,39) and (136.5,47) .. (134,50.5) .. controls (131.5,54) and (97.5,68.5) .. (91,68.5) .. controls (84.5,68.5) and (62.5,54.5) .. (65,50.5) .. controls (67.5,46.5) and (107.5,39) .. (111,39) -- cycle ;
%Shape: Polygon Curved [id:ds5992111128710786] 
\draw  [draw opacity=0][fill={rgb, 255:red, 74; green, 144; blue, 226 }  ,fill opacity=0.25 ] (117.5,147) .. controls (120.5,147) and (130.5,152) .. (130,156) .. controls (129.5,160) and (117.5,160.5) .. (114,160.5) .. controls (110.5,160.5) and (102,155) .. (103.5,152.5) .. controls (105,150) and (114.5,147) .. (117.5,147) -- cycle ;
%Straight Lines [id:da9464023698506874] 
\draw [color={rgb, 255:red, 128; green, 128; blue, 128 }  ,draw opacity=1 ]   (57.5,145.5) -- (57.5,105) ;
\draw [shift={(57.5,103)}, rotate = 90] [color={rgb, 255:red, 128; green, 128; blue, 128 }  ,draw opacity=1 ][line width=0.75]    (10.93,-3.29) .. controls (6.95,-1.4) and (3.31,-0.3) .. (0,0) .. controls (3.31,0.3) and (6.95,1.4) .. (10.93,3.29)   ;

% Text Node
\draw (93.52,50.7) node    {$\mathscr{A}$};
% Text Node
\draw (46.22,123.2) node  [color={rgb, 255:red, 128; green, 128; blue, 128 }  ,opacity=1 ]  {$v$};

\end{tikzpicture}

    \caption{An expanding null geodesic congruence. The area spanned by nearby geodesics provides a measure of proximity among the light rays.}
\end{figure}

The focusing theorem in GR can be proved by using the null-null component of the Einstein equation $G_{vv} = R_{vv} = 8 \pi G T_{vv}$ where $G_{ab} = R_{ab} - \frac{1}{2} g_{ab} R$ is the Einstein tensor, $G$ is Newton's constant, $T_{ab}$ is the stress-energy tensor; and the null energy condition (NEC) $T_{vv} \geq 0$. After the substitution of the Einstein equation, we now call the Raychaudhuri equation a \emph{focusing equation} as it describes how the focusing of null geodesics is sourced dynamically, and its r.h.s. is non-positive:
\begin{equation}
    \partial_v \theta = - \frac{1}{D-2} \theta^2 - \sigma_{ab} \sigma^{ab} - 8 \pi G T_{vv} \leq 0.
\end{equation}
Then, the expansion is always non-increasing. Hence, light rays always tend to focus. We observe that the NEC is essential because, in the limit where the self-focusing terms $\theta^2$ and $\sigma^2$ are negligible, a positive energy density is crucial to ensure the convergence of light beams. 

The focusing equation exhibits a curious feature when evaluated on a horizon: it encodes the thermodynamics of the horizon! This is easy to see upon the identification that the horizon's entropy (Bekenstein-Hawking entropy) is proportional to its cross-section area:
\begin{equation}
 S_{\text{BH}} = \frac{A}{4 G}.
\end{equation}
Thus, the expansion can be interpreted as the rate of change in entropy density across a null surface. Moreover, the focusing equation and the focusing theorem imply both the first and second laws of horizon thermodynamics.

Start with a bifurcate stationary\footnote{An entropy is better defined when it is near thermal equilibrium. Bifurcate Killing horizons enjoy a zeroth law that the horizon temperature is constant. \cite{Bardeen:1973gs,Jacobson:1995uq,Ghosh:2020dkk,Bhattacharyya:2022nqa,Davies:2024fut}} horizon $\mathcal{H}$ with Killing vector $\xi$. The null generators of $\mathcal{H}$ are all parallel to each other, i.e., $\theta = 0$ and $\sigma = 0$. Perturb it by an infinitesimal matter source with energy density $T_{ab}$ that satisfies NEC, assuming a teleological boundary condition such that the horizon settles down to stationarity at far future $v \to \infty$. Integrating the focusing equation on a cross-section $\mathcal{C}$ of the horizon, we have 
\begin{equation}
    \partial_v^2 S_{\text{BH}}[\mathcal{C}] = - 2 \pi \int_\mathcal{C} T_{vv} \dd{A}.
\end{equation} 
Inverting the second $v$-derivative by integration, we reach a physical version of the first law \cite{Wald:1995yp}
\begin{equation}
 \frac{\kappa}{2 \pi} \Delta S_{\text{BH}} = \int_{0}^{\infty} \dd{v} \int \dd{A} T_{ab} k^a \xi^b = \Delta M - \Omega_{\mathcal{H}} \Delta J
\end{equation}
where $\Delta$ labels the change between the future infinity and the bifurcation surface $\mathcal{B}$, $M$ and $J$ are the mass and angular momentum of the infalling matter source, and $\Omega_\mathcal{H}$ is the horizon angular speed. On the other hand, the second law is given by the focusing theorem together with the teleological boundary condition:
\begin{equation}
    \partial_v^2 S_\text{BH} \leq 0 \quad \text{and} \quad \partial_v S_\text{BH} \to 0 \text{ as } v \to 0 \qquad \Rightarrow \qquad \partial_v S_\text{BH} \geq 0.
\end{equation}

Conversely, instead of using the dynamics as an input, one can use the thermodynamic laws and the Raychaudhuri equation of the horizon to determine the Einstein equation as an equation of state \cite{Jacobson:1995ab}. Either way, we have seen the critical roles that the Raychaudhuri/focusing equations play in decoding the thermodynamics of Killing horizons.

\subsection{Generalised focusing equation in diffeomorphism-invariant theories}
The Raychaudhuri/focusing equations contain significant information about the entropy of spacetime and associated thermodynamic laws. Its generalisation to include quantum corrections to GR could offer us vital clues for decoding the microstates of quantum gravity. This has been proven successful in many ways. For the scenario of semi-classical gravity that couples to quantum matter fields, the expansion can be generalised to a quantum expansion, and a Quantum Focusing Conjecture stating that quantum expansion cannot increase has been proposed \cite{Bousso:2015mna}. Accordingly, the Bekenstein-Hawking entropy is promoted to the generalised entropy \cite{Bekenstein:1973ur,Hawking:1975vcx} which adds the entanglement entropy of the quantum fields exterior to the horizon alongside $A/4G$, and a quantum version of the singularity theorem can be proven given the generalised second law holds \cite{Wall:2010jtc}.

Here, we take a different route: we consider the higher-derivative modification to GR as a result of quantum loop and/or stringy corrections at the level of classical low-energy effective theories of quantum gravity. Previously, the Raychaudhuri/focusing equations were gradually generalised to arbitrary diffeomorphism-invariant theories of gravity non-minimally coupled to scalar and vector fields \cite{Kolekar:2012tq,Sarkar:2013swa,Bhattacharjee:2015yaa,Wall:2015raa,Bhattacharjee:2015qaa,Bhattacharyya:2016xfs,Bhattacharya:2019qal,Bhattacharyya:2021jhr,Biswas:2022grc,Deo:2023vvb,Wall:2024lbd}. Due to the extreme generality of the class of theories in consideration, the analyses are only carried out on bifurcate Killing horizons where a generalised expansion is well-defined to the linear order of dynamical perturbations.\footnote{Additional assumptions or constraints are required to study non-linear perturbations. For instance, \cite{Hollands:2022fkn} restricted attention to effective field theories whose parameters are controlled by a characteristic length scale $\ell$ in order to study second-order perturbations.} Below, we outline the generalisation to the case of gravity coupled to scalar and/or vector fields.

Among all diffeomorphism-invariant theories, General Relativity is a special theory---its geometry and dynamics are compatible in the sense that positive energy always ensures the focusing of light rays. Now, consider an arbitrary theory of gravity---presumably GR plus higher-derivative corrections, the gravitational equation of motion becomes
\begin{equation}
 E_{ab} \equiv \frac{1}{8 \pi G} G_{ab} + H_{ab} = T_{ab}
\end{equation}
where $H_{ab}$ is the correction terms to Einstein equation. As a convention, we have moved the $8 \pi G$ factor before the Einstein tensor. Now, the geometry and dynamics are not compatible: the focusing equation becomes
\begin{equation}
    \partial_v \theta = - \frac{1}{D - 2} \theta^2 - \sigma_{ab} \sigma^{ab} - 8 \pi G T_{vv} + 8 \pi G H_{vv}.
\end{equation}
In general, the correction term $H_{vv}$ cannot be guaranteed with a definite sign. The focusing theorem of light rays is violated even if $T_{ab}$ obeys NEC. The null geodesics no longer focus in the sense of a shrinking area element. This is expected also from the perspective of entropy: the area density is no longer the entropy density. 

We need to invent a new notion of ``focusing'' of light rays. This essentially requires a generalised measure of proximity between null geodesics beyond the geometrical area density. For null geodesics to focus with respect to this new measure, the associated \emph{generalised expansion} $\Theta$ (along with a generalised shear $\Sigma_{ab}$) must satisfy an updated version of Raychaudhuri/focusing equation by absorbing the contributions from $H_{vv}$:
\begin{equation}
    \begin{split}
        \partial_v \Theta & \overset{?}{=} - \frac{1}{D-2} \Theta^2 - \Sigma_{ab} \Sigma^{ab} - 2 \pi E_{vv}\\
        & = - \frac{1}{D-2} \Theta^2 - \Sigma_{ab} \Sigma^{ab} - 2 \pi T_{vv} \leq 0
    \end{split}
\end{equation}
where $2 \pi$ is merely a convention, which results in non-positive $\partial_v \Theta$ when NEC is satisfied. In other words, to find such a valid generalised expansion, we need to analyse the structure of the equation of motion tensor $E_{ab}$ for general diffeomorphism-invariant theory and show its null-null component can be written as a total $v$-derivative term minus a quadratic piece. In general, this is very hard, given the arbitrariness of the theories and the non-linearity of the equations. Therefore, we reduce our attention to Killing horizons and their linear dynamical perturbations only to get tangible results. The background Killing symmetry is powerful enough to resolve the structure of $E_{vv}$ \emph{on the horizon}, and it can be proven that a generalised expansion $\Theta$ with desired properties is admitted in this regime. This scope of consideration is also satisfactory enough to extract the dynamical horizon entropy and to prove the first and second laws of thermodynamics up to the linear order of perturbation. 

The background Killing symmetry enables a powerful \emph{boost weight analysis}, which determines the structure of $E_{vv}$ in some Gaussian null coordinates (GNC) $u,v,x^i$ where $v$ is the affine null coordinate along the horizon generators, $u$ is the other null direction, and $x^i$ are codimension-2 spatial coordinates. It is shown, for gravity non-minimally coupled to scalar and/or vector fields, that \cite{Bhattacharya:2019qal,Bhattacharyya:2021jhr,Biswas:2022grc,Deo:2023vvb,Wall:2024lbd}
\begin{equation}
 E_{vv} \fheq - \frac{1}{2 \pi} \partial_v \Theta \label{eq:GLRE}
\end{equation}
on the future horizon $\mathcal{H}^+$, where the generalised expansion $\Theta$ can be read off. Its structure is further shown as
\begin{equation}
    \Theta = \partial_v \varsigma + D_i J^i
\end{equation}
where $\varsigma$ is an entropy density, $D_i$ is the intrinsic covariant derivative in $x^i$ directions, and $J^i$ is an entropy current. In other words, the generalised expansion is the divergence of an \emph{entropy density-current vector} $(\varsigma, J^i)$ on the horizon, which is similar to the l.h.s.~of a continuity equation. Importantly, this entropy density-current is precisely the new measure of the proximity of nearby null geodesics. The structure of $E_{vv}$ (eq.~\eqref{eq:GLRE}) gives us the \emph{generalised linear Raychaudhuri equation}, and it is consistent with the fact that $\Theta^2$ and $\Sigma_{ab} \Sigma^{ab}$ terms could be ignored at the linear order of perturbations to Killing horizons. When the equations of motion are applied, we have a \emph{linearised focusing theorem} 
\begin{equation}
    \partial_v \Theta = - 2 \pi T_{vv} \leq 0 \label{eq:GLFE}
\end{equation}
given that NEC holds.

\begin{figure}[H]
    \centering

    \tikzset{every picture/.style={line width=0.75pt}} %set default line width to 0.75pt        

    \begin{tikzpicture}[x=0.75pt,y=0.75pt,yscale=-1,xscale=1]
    
    %Curve Lines [id:da891090019562002] 
    \draw    (109.5,72.5) .. controls (109.94,77.17) and (117.34,88.11) .. (112,100) .. controls (106.66,111.89) and (85.94,130.4) .. (84.5,141.5) .. controls (83.06,152.6) and (99.21,168.43) .. (98.5,176.5) ;
    \draw [shift={(98.5,176.5)}, rotate = 95.02] [color={rgb, 255:red, 0; green, 0; blue, 0 }  ][fill={rgb, 255:red, 0; green, 0; blue, 0 }  ][line width=0.75]      (0, 0) circle [x radius= 1.34, y radius= 1.34]   ;
    \draw [shift={(112.88,83.4)}, rotate = 80.34] [fill={rgb, 255:red, 0; green, 0; blue, 0 }  ][line width=0.08]  [draw opacity=0] (3.57,-1.72) -- (0,0) -- (3.57,1.72) -- cycle    ;
    \draw [shift={(99.15,118.12)}, rotate = 128.81] [fill={rgb, 255:red, 0; green, 0; blue, 0 }  ][line width=0.08]  [draw opacity=0] (3.57,-1.72) -- (0,0) -- (3.57,1.72) -- cycle    ;
    \draw [shift={(89.33,156.87)}, rotate = 59.64] [fill={rgb, 255:red, 0; green, 0; blue, 0 }  ][line width=0.08]  [draw opacity=0] (3.57,-1.72) -- (0,0) -- (3.57,1.72) -- cycle    ;
    \draw [shift={(109.5,72.5)}, rotate = 84.59] [color={rgb, 255:red, 0; green, 0; blue, 0 }  ][fill={rgb, 255:red, 0; green, 0; blue, 0 }  ][line width=0.75]      (0, 0) circle [x radius= 1.34, y radius= 1.34]   ;
    %Curve Lines [id:da4913977628048276] 
    \draw    (120.5,81) .. controls (121.32,98.56) and (112.07,114.46) .. (114.5,127.5) .. controls (116.93,140.54) and (127.34,142.1) .. (130,150.5) .. controls (132.66,158.9) and (128.7,173.14) .. (134,180.5) ;
    \draw [shift={(134,180.5)}, rotate = 54.21] [color={rgb, 255:red, 0; green, 0; blue, 0 }  ][fill={rgb, 255:red, 0; green, 0; blue, 0 }  ][line width=0.75]      (0, 0) circle [x radius= 1.34, y radius= 1.34]   ;
    \draw [shift={(117.93,101.68)}, rotate = 103.62] [fill={rgb, 255:red, 0; green, 0; blue, 0 }  ][line width=0.08]  [draw opacity=0] (3.57,-1.72) -- (0,0) -- (3.57,1.72) -- cycle    ;
    \draw [shift={(119.22,137.65)}, rotate = 45.39] [fill={rgb, 255:red, 0; green, 0; blue, 0 }  ][line width=0.08]  [draw opacity=0] (3.57,-1.72) -- (0,0) -- (3.57,1.72) -- cycle    ;
    \draw [shift={(131.02,162.9)}, rotate = 89.72] [fill={rgb, 255:red, 0; green, 0; blue, 0 }  ][line width=0.08]  [draw opacity=0] (3.57,-1.72) -- (0,0) -- (3.57,1.72) -- cycle    ;
    \draw [shift={(120.5,81)}, rotate = 87.34] [color={rgb, 255:red, 0; green, 0; blue, 0 }  ][fill={rgb, 255:red, 0; green, 0; blue, 0 }  ][line width=0.75]      (0, 0) circle [x radius= 1.34, y radius= 1.34]   ;
    %Curve Lines [id:da4524631624718718] 
    \draw    (133,66.5) .. controls (137.31,93.8) and (126.91,82.89) .. (127.5,101.5) .. controls (128.09,120.11) and (139.61,124.44) .. (143,137) .. controls (146.39,149.56) and (137.69,160.16) .. (137.5,167) ;
    \draw [shift={(137.5,167)}, rotate = 91.57] [color={rgb, 255:red, 0; green, 0; blue, 0 }  ][fill={rgb, 255:red, 0; green, 0; blue, 0 }  ][line width=0.75]      (0, 0) circle [x radius= 1.34, y radius= 1.34]   ;
    \draw [shift={(133.72,81.93)}, rotate = 111.55] [fill={rgb, 255:red, 0; green, 0; blue, 0 }  ][line width=0.08]  [draw opacity=0] (3.57,-1.72) -- (0,0) -- (3.57,1.72) -- cycle    ;
    \draw [shift={(131.77,117.6)}, rotate = 57.53] [fill={rgb, 255:red, 0; green, 0; blue, 0 }  ][line width=0.08]  [draw opacity=0] (3.57,-1.72) -- (0,0) -- (3.57,1.72) -- cycle    ;
    \draw [shift={(142.96,149.56)}, rotate = 107.28] [fill={rgb, 255:red, 0; green, 0; blue, 0 }  ][line width=0.08]  [draw opacity=0] (3.57,-1.72) -- (0,0) -- (3.57,1.72) -- cycle    ;
    \draw [shift={(133,66.5)}, rotate = 81.03] [color={rgb, 255:red, 0; green, 0; blue, 0 }  ][fill={rgb, 255:red, 0; green, 0; blue, 0 }  ][line width=0.75]      (0, 0) circle [x radius= 1.34, y radius= 1.34]   ;
    %Curve Lines [id:da682336047621064] 
    \draw    (142,72) .. controls (143.89,84.34) and (161.12,78.12) .. (159,90.5) .. controls (156.88,102.88) and (147.88,128.35) .. (150.5,137.5) .. controls (153.12,146.65) and (180.33,156.18) .. (175.5,170.5) ;
    \draw [shift={(175.5,170.5)}, rotate = 108.66] [color={rgb, 255:red, 0; green, 0; blue, 0 }  ][fill={rgb, 255:red, 0; green, 0; blue, 0 }  ][line width=0.75]      (0, 0) circle [x radius= 1.34, y radius= 1.34]   ;
    \draw [shift={(148.64,79.96)}, rotate = 20.91] [fill={rgb, 255:red, 0; green, 0; blue, 0 }  ][line width=0.08]  [draw opacity=0] (3.57,-1.72) -- (0,0) -- (3.57,1.72) -- cycle    ;
    \draw [shift={(153.99,111.12)}, rotate = 103.77] [fill={rgb, 255:red, 0; green, 0; blue, 0 }  ][line width=0.08]  [draw opacity=0] (3.57,-1.72) -- (0,0) -- (3.57,1.72) -- cycle    ;
    \draw [shift={(164.02,150.64)}, rotate = 39.8] [fill={rgb, 255:red, 0; green, 0; blue, 0 }  ][line width=0.08]  [draw opacity=0] (3.57,-1.72) -- (0,0) -- (3.57,1.72) -- cycle    ;
    \draw [shift={(142,72)}, rotate = 81.31] [color={rgb, 255:red, 0; green, 0; blue, 0 }  ][fill={rgb, 255:red, 0; green, 0; blue, 0 }  ][line width=0.75]      (0, 0) circle [x radius= 1.34, y radius= 1.34]   ;
    %Shape: Parallelogram [id:dp5303546612777787] 
    \draw   (84.8,155) -- (218.5,155) -- (161.2,195) -- (27.5,195) -- cycle ;
    %Shape: Parallelogram [id:dp18331730715948003] 
    \draw   (97.65,49.5) -- (213.5,49.5) -- (164.85,95.5) -- (49,95.5) -- cycle ;
    %Shape: Polygon Curved [id:ds18786103709008062] 
    \draw  [draw opacity=0][fill={rgb, 255:red, 74; green, 144; blue, 226 }  ,fill opacity=0.25 ] (133,66.5) .. controls (136.5,66.5) and (144.5,68.5) .. (142,72) .. controls (139.5,75.5) and (127,81) .. (120.5,81) .. controls (114,81) and (107,76.5) .. (109.5,72.5) .. controls (112,68.5) and (129.5,66.5) .. (133,66.5) -- cycle ;
    %Shape: Polygon Curved [id:ds11401510415155003] 
    \draw  [draw opacity=0][fill={rgb, 255:red, 74; green, 144; blue, 226 }  ,fill opacity=0.25 ] (137.5,167) .. controls (140.5,167) and (176,166.5) .. (175.5,170.5) .. controls (175,174.5) and (137.5,180.5) .. (134,180.5) .. controls (130.5,180.5) and (97,179) .. (98.5,176.5) .. controls (100,174) and (134.5,167) .. (137.5,167) -- cycle ;
    %Straight Lines [id:da989679195212787] 
    \draw [color={rgb, 255:red, 128; green, 128; blue, 128 }  ,draw opacity=1 ]   (58,153.5) -- (58,113) ;
    \draw [shift={(58,111)}, rotate = 90] [color={rgb, 255:red, 128; green, 128; blue, 128 }  ,draw opacity=1 ][line width=0.75]    (10.93,-3.29) .. controls (6.95,-1.4) and (3.31,-0.3) .. (0,0) .. controls (3.31,0.3) and (6.95,1.4) .. (10.93,3.29)   ;
    %Curve Lines [id:da9061044379856682] 
    \draw    (301,71) .. controls (305.5,118.5) and (309,139.5) .. (339.5,173) ;
    \draw [shift={(339.5,173)}, rotate = 47.68] [color={rgb, 255:red, 0; green, 0; blue, 0 }  ][fill={rgb, 255:red, 0; green, 0; blue, 0 }  ][line width=0.75]      (0, 0) circle [x radius= 1.34, y radius= 1.34]   ;
    \draw [shift={(309.3,123.52)}, rotate = 71.65] [fill={rgb, 255:red, 0; green, 0; blue, 0 }  ][line width=0.08]  [draw opacity=0] (3.57,-1.72) -- (0,0) -- (3.57,1.72) -- cycle    ;
    \draw [shift={(301,71)}, rotate = 84.59] [color={rgb, 255:red, 0; green, 0; blue, 0 }  ][fill={rgb, 255:red, 0; green, 0; blue, 0 }  ][line width=0.75]      (0, 0) circle [x radius= 1.34, y radius= 1.34]   ;
    %Curve Lines [id:da8196317100820265] 
    \draw    (327,89) .. controls (329,132) and (334.5,159.5) .. (350,181) ;
    \draw [shift={(350,181)}, rotate = 54.21] [color={rgb, 255:red, 0; green, 0; blue, 0 }  ][fill={rgb, 255:red, 0; green, 0; blue, 0 }  ][line width=0.75]      (0, 0) circle [x radius= 1.34, y radius= 1.34]   ;
    \draw [shift={(331.58,134.25)}, rotate = 78.67] [fill={rgb, 255:red, 0; green, 0; blue, 0 }  ][line width=0.08]  [draw opacity=0] (3.57,-1.72) -- (0,0) -- (3.57,1.72) -- cycle    ;
    \draw [shift={(327,89)}, rotate = 87.34] [color={rgb, 255:red, 0; green, 0; blue, 0 }  ][fill={rgb, 255:red, 0; green, 0; blue, 0 }  ][line width=0.75]      (0, 0) circle [x radius= 1.34, y radius= 1.34]   ;
    %Curve Lines [id:da5882430369239862] 
    \draw    (347,59.5) .. controls (354.5,107) and (354.5,131) .. (353.5,167.5) ;
    \draw [shift={(353.5,167.5)}, rotate = 91.57] [color={rgb, 255:red, 0; green, 0; blue, 0 }  ][fill={rgb, 255:red, 0; green, 0; blue, 0 }  ][line width=0.75]      (0, 0) circle [x radius= 1.34, y radius= 1.34]   ;
    \draw [shift={(353.03,110.6)}, rotate = 86.62] [fill={rgb, 255:red, 0; green, 0; blue, 0 }  ][line width=0.08]  [draw opacity=0] (3.57,-1.72) -- (0,0) -- (3.57,1.72) -- cycle    ;
    \draw [shift={(347,59.5)}, rotate = 81.03] [color={rgb, 255:red, 0; green, 0; blue, 0 }  ][fill={rgb, 255:red, 0; green, 0; blue, 0 }  ][line width=0.75]      (0, 0) circle [x radius= 1.34, y radius= 1.34]   ;
    %Curve Lines [id:da8827984936860775] 
    \draw    (370,71) .. controls (375.5,107) and (379,138) .. (366,176.5) ;
    \draw [shift={(366,176.5)}, rotate = 108.66] [color={rgb, 255:red, 0; green, 0; blue, 0 }  ][fill={rgb, 255:red, 0; green, 0; blue, 0 }  ][line width=0.75]      (0, 0) circle [x radius= 1.34, y radius= 1.34]   ;
    \draw [shift={(374.94,121.28)}, rotate = 90.69] [fill={rgb, 255:red, 0; green, 0; blue, 0 }  ][line width=0.08]  [draw opacity=0] (3.57,-1.72) -- (0,0) -- (3.57,1.72) -- cycle    ;
    \draw [shift={(370,71)}, rotate = 81.31] [color={rgb, 255:red, 0; green, 0; blue, 0 }  ][fill={rgb, 255:red, 0; green, 0; blue, 0 }  ][line width=0.75]      (0, 0) circle [x radius= 1.34, y radius= 1.34]   ;
    %Shape: Parallelogram [id:dp568730223430032] 
    \draw   (330.7,155.5) -- (393,155.5) -- (366.3,195.5) -- (304,195.5) -- cycle ;
    %Shape: Parallelogram [id:dp7950721662177334] 
    \draw   (307,50) -- (407,50) -- (365,99) -- (265,99) -- cycle ;
    %Shape: Polygon Curved [id:ds27301599083484773] 
    \draw  [draw opacity=0][fill={rgb, 255:red, 208; green, 2; blue, 27 }  ,fill opacity=0.25 ] (347,59.5) .. controls (350.5,59.5) and (372.5,67.5) .. (370,71) .. controls (367.5,74.5) and (333.5,89) .. (327,89) .. controls (320.5,89) and (298.5,75) .. (301,71) .. controls (303.5,67) and (343.5,59.5) .. (347,59.5) -- cycle ;
    %Shape: Polygon Curved [id:ds2694707860100036] 
    \draw  [draw opacity=0][fill={rgb, 255:red, 208; green, 2; blue, 27 }  ,fill opacity=0.25 ] (353.5,167.5) .. controls (356.5,167.5) and (366.5,172.5) .. (366,176.5) .. controls (365.5,180.5) and (353.5,181) .. (350,181) .. controls (346.5,181) and (338,175.5) .. (339.5,173) .. controls (341,170.5) and (350.5,167.5) .. (353.5,167.5) -- cycle ;
    %Straight Lines [id:da8493468577337531] 
    \draw [color={rgb, 255:red, 128; green, 128; blue, 128 }  ,draw opacity=1 ]   (293.5,166) -- (293.5,125.5) ;
    \draw [shift={(293.5,123.5)}, rotate = 90] [color={rgb, 255:red, 128; green, 128; blue, 128 }  ,draw opacity=1 ][line width=0.75]    (10.93,-3.29) .. controls (6.95,-1.4) and (3.31,-0.3) .. (0,0) .. controls (3.31,0.3) and (6.95,1.4) .. (10.93,3.29)   ;
    
    % Text Node
    \draw (122.52,72.2) node    {$\mathscr{A}$};
    % Text Node
    \draw (46.72,131.2) node  [color={rgb, 255:red, 128; green, 128; blue, 128 }  ,opacity=1 ]  {$v$};
    % Text Node
    \draw (334,74.5) node    {$\varsigma, J^i $};
    % Text Node
    \draw (282.22,143.7) node  [color={rgb, 255:red, 128; green, 128; blue, 128 }  ,opacity=1 ]  {$v$};
    % Text Node
    \draw (122.46,226.5) node   [align=left] {\begin{minipage}[lt]{116.12pt}\setlength\topsep{0pt}
    \begin{center}
    Light rays in area density\\(exaggerated)
    \end{center}
    
    \end{minipage}};
    % Text Node
    \draw (345.96,223) node   [align=left] {\begin{minipage}[lt]{129.73pt}\setlength\topsep{0pt}
    \begin{center}
    Light rays in entropy density-current
    \end{center}
    
    \end{minipage}};

    \end{tikzpicture}
    
\caption{In general theories of gravity, light rays may anti-focus when the proximity is measured by the area density. Focusing is restored if we use the entropy density-current instead to measure how close the null geodesics are.}
\end{figure}

Subsequently, we can use the above linearised focusing equation to study the entropy of the horizon. For horizons with compact cross-sections,\footnote{For non-compact horizons, the entropy current can leak through the boundary of horizon slices, and it is no longer a closed system. The entropy can decrease in that case, and the second law is no longer true.} the Wall entropy (also known as the increasing entropy) \cite{Wall:2015raa}---a dynamical generalisation of the Wald entropy \cite{Wald:1993nt}---was originally defined by 
\begin{equation}
    \partial_v^2 S_{\text{Wall}}[\mathcal{C}] = - 2 \pi \int_\mathcal{C} T_{vv} \dd{A}  \label{eq:S-Wall-def}
\end{equation}
for some horizon slice $\mathcal{C}$. Using the generalised focusing equation \eqref{eq:GLFE}, the Wall entropy can be extracted as 
\begin{equation}
    S_\text{Wall}[\mathcal{C}] = \int_\mathcal{C} \varsigma \dd{A},  \label{eq:S-Wall}
\end{equation}
and it satisfies both the first and the linearised second law (assuming teleological boundary condition again)
\begin{equation}
 \frac{\kappa}{2 \pi} \Delta S_\text{Wall} = \Delta M - \Omega_\mathcal{H} \Delta J, \qquad \partial_v S_\text{Wall} \geq 0.
\end{equation}

To give an impression, in $f(\text{Riemann})$ gravity, the Wall entropy reads \cite{Wall:2015raa}
\begin{equation}
 S_\text{Wall} = - 2 \pi \int_\mathcal{C} \dd{A} \left(4 \pdv{L}{R_{uvuv}} + 16 \pdv{L}{R_{uiuj}}{R_{vkvl}} \bar K_{ij} K_{kl}\right)
\end{equation}
where the first term is the Wald entropy, and the second term is a dynamical correction with $\bar K_{ij}, K_{kl}$ extrinsic curvatures in $u,v$-directions, respectively.

\begin{figure}[H]
    \centering

\tikzset{every picture/.style={line width=0.75pt}} %set default line width to 0.75pt        

\begin{tikzpicture}[x=0.75pt,y=0.75pt,yscale=-1,xscale=1]

%Curve Lines [id:da9220880038006649] 
\draw    (119.04,66.17) .. controls (64.51,161.67) and (67.34,179.17) .. (66.33,238.17) ;
%Curve Lines [id:da8027480109496379] 
\draw [color={rgb, 255:red, 74; green, 144; blue, 226 }  ,draw opacity=1 ]   (134.22,67.07) .. controls (88.38,149.07) and (74.2,179.27) .. (73.19,238.27) ;
\draw [shift={(94.66,142.53)}, rotate = 113.77] [color={rgb, 255:red, 74; green, 144; blue, 226 }  ,draw opacity=1 ][line width=0.75]    (10.93,-3.29) .. controls (6.95,-1.4) and (3.31,-0.3) .. (0,0) .. controls (3.31,0.3) and (6.95,1.4) .. (10.93,3.29)   ;
%Curve Lines [id:da3702570792596256] 
\draw [color={rgb, 255:red, 74; green, 144; blue, 226 }  ,draw opacity=1 ]   (149.63,67.47) .. controls (114.64,133.07) and (87.58,178.67) .. (78.6,238.67) ;
\draw [shift={(109.31,144.55)}, rotate = 115.3] [color={rgb, 255:red, 74; green, 144; blue, 226 }  ,draw opacity=1 ][line width=0.75]    (10.93,-3.29) .. controls (6.95,-1.4) and (3.31,-0.3) .. (0,0) .. controls (3.31,0.3) and (6.95,1.4) .. (10.93,3.29)   ;
%Straight Lines [id:da8761594831057338] 
\draw [color={rgb, 255:red, 74; green, 144; blue, 226 }  ,draw opacity=1 ]   (178.41,69.67) -- (88.84,239.46) ;
\draw [shift={(136.89,148.37)}, rotate = 117.81] [color={rgb, 255:red, 74; green, 144; blue, 226 }  ,draw opacity=1 ][line width=0.75]    (10.93,-3.29) .. controls (6.95,-1.4) and (3.31,-0.3) .. (0,0) .. controls (3.31,0.3) and (6.95,1.4) .. (10.93,3.29)   ;
%Curve Lines [id:da8953475187153672] 
\draw [color={rgb, 255:red, 74; green, 144; blue, 226 }  ,draw opacity=1 ]   (163.5,69.47) .. controls (124.41,143.87) and (95.48,191.97) .. (83.83,239.47) ;
\draw [shift={(122.37,146.9)}, rotate = 117.35] [color={rgb, 255:red, 74; green, 144; blue, 226 }  ,draw opacity=1 ][line width=0.75]    (10.93,-3.29) .. controls (6.95,-1.4) and (3.31,-0.3) .. (0,0) .. controls (3.31,0.3) and (6.95,1.4) .. (10.93,3.29)   ;
%Curve Lines [id:da71303651490392] 
\draw    (237.98,73.49) .. controls (190.97,168.17) and (170.74,185.49) .. (111.35,240.75) ;
%Curve Lines [id:da4158415226556589] 
\draw [color={rgb, 255:red, 74; green, 144; blue, 226 }  ,draw opacity=1 ]   (223.61,72.67) .. controls (180.51,155.77) and (163.64,185.01) .. (104.25,240.27) ;
\draw [shift={(176.59,157.12)}, rotate = 123.4] [color={rgb, 255:red, 74; green, 144; blue, 226 }  ,draw opacity=1 ][line width=0.75]    (10.93,-3.29) .. controls (6.95,-1.4) and (3.31,-0.3) .. (0,0) .. controls (3.31,0.3) and (6.95,1.4) .. (10.93,3.29)   ;
%Curve Lines [id:da6189987969332995] 
\draw [color={rgb, 255:red, 74; green, 144; blue, 226 }  ,draw opacity=1 ]   (208.38,71.87) .. controls (171.66,139.47) and (152.2,182.69) .. (99.24,239.87) ;
\draw [shift={(163.25,154.21)}, rotate = 121.03] [color={rgb, 255:red, 74; green, 144; blue, 226 }  ,draw opacity=1 ][line width=0.75]    (10.93,-3.29) .. controls (6.95,-1.4) and (3.31,-0.3) .. (0,0) .. controls (3.31,0.3) and (6.95,1.4) .. (10.93,3.29)   ;
%Curve Lines [id:da9295288445278203] 
\draw [color={rgb, 255:red, 74; green, 144; blue, 226 }  ,draw opacity=1 ]   (193.13,70.27) .. controls (157.71,135.07) and (131.33,194.04) .. (93.64,239.87) ;
\draw [shift={(149.89,151.79)}, rotate = 118.44] [color={rgb, 255:red, 74; green, 144; blue, 226 }  ,draw opacity=1 ][line width=0.75]    (10.93,-3.29) .. controls (6.95,-1.4) and (3.31,-0.3) .. (0,0) .. controls (3.31,0.3) and (6.95,1.4) .. (10.93,3.29)   ;
%Straight Lines [id:da12476294322437953] 
\draw    (119.04,66.17) -- (237.98,73.49) ;
%Straight Lines [id:da6689225850770366] 
\draw    (66.33,238.17) -- (111.35,240.75) ;
%Straight Lines [id:da6276144510359045] 
\draw    (44.87,131.6) -- (66.09,90.11) ;
\draw [shift={(67,88.33)}, rotate = 117.09] [color={rgb, 255:red, 0; green, 0; blue, 0 }  ][line width=0.75]    (10.93,-3.29) .. controls (6.95,-1.4) and (3.31,-0.3) .. (0,0) .. controls (3.31,0.3) and (6.95,1.4) .. (10.93,3.29)   ;

% Text Node
\draw (97.64,68.8) node    {$\mathcal{H}$};
% Text Node
\draw (285.31,76.79) node    {$\partial _{v} S_{\text{Wall}}\rightarrow 0$};
% Text Node
\draw (45.56,105.2) node    {$v$};
% Text Node
\draw (176.68,56.52) node    {$S_{\text{Wall}}( v_{2})$};
% Text Node
\draw (87.68,253.35) node    {$S_{\text{Wall}}( v_{1})$};

\end{tikzpicture}
\caption{In general theories of gravity, horizons are expanding (to the first order of dynamical perturbation) if their sizes are quantified by their Wall entropies.}
\end{figure}

\subsection{Gravitational focusing, horizon entropy, and higher-spin fields}
The previous works have not touched on some important elements---spin $s \geq 2$ bosonic tensor fields excluding the metric. These include massive gravitons and the conventional $s\geq 3$ higher-spin fields.\footnote{The rank $s\geq 2$ bosonic tensor fields in our consideration also include $p$-form gauge fields. However, they are effectively spin-1 in the gravitational focusing scenario, as will be discussed in Section \ref{sec:discussion}.} Generally speaking, they are rather exotic objects, which often lead to puzzles and inconsistencies, and the requirement of physical consistency imposes stringent constraints on the types of interactions. However, the search for physically consistent spin $s \geq 2$ fields is well motivated both theoretically and phenomenologically. On one hand, they are crucial for a better understanding of the ultimate theory of quantum gravity. As an important example, String Theory predicts an infinite tower of massive higher-spin fields, including massive spin-2 excitations in the open string spectrum. These become massless in the tensionless limit of strings. On the other hand, there exist massive higher-spin composite particles in nature (e.g., hadrons and nuclei); also, massive gravitons could offer important clues for the accelerated expansion of our universe and the cosmological constant problem.

In this paper, we study these spin $s \geq 2$ fields from the angle of gravitational focusing and horizon thermodynamics. The main objective of our work is to demonstrate how pathologies arise in the guise of violation of the focusing theorem when some arbitrary $s \geq 2$ fields are present. Before turning to this, we give a very brief introduction to massive gravity and higher-spin fields and their associated problems. (See e.g.~\cite{deRham:2014zqa,Bekaert:2010hw} for detailed reviews on these subjects.)

Extending gravity by giving masses to gravitons was first considered by Fierz and Pauli \cite{Fierz:1939ix}. They switched on a mass term for the graviton $\gamma_{ab}$ alongside the linearised Einstein-Hilbert Lagrangian:
\begin{equation}
    L_{\text{FP}} = - \frac{1}{8}m^2 (\gamma_{ab}\gamma^{ab} - \gamma^2)
\end{equation}
where $m$ is the mass and $\gamma = g^{ab} \gamma_{ab}$ is the trace of the massive graviton on a background metric $g_{ab}$. This is the natural generalisation of the massive scalar and vector fields. However, it is plagued by two major challenges. The first one is the van Dam-Veltman-Zakharov (vDVZ) discontinuity \cite{vanDam:1970vg}: in the massless limit, massive gravity does not smoothly recover GR predictions. The extra scalar mode in the massive graviton introduces additional attraction for massive matter fields compared with light-like fields. As a result, in the massless limit of massive gravity, the bending of light is only 3/4 of that predicted in GR. The second problem arises when constructing a non-linear generalisation of Fierz-Pauli: the resulting theory can have a Boulware-Deser (BD) ghost \cite{Boulware:1972yco}. Subsequently, the first problem was cured by Vainshtein's mechanism \cite{Vainshtein:1972sx} which demonstrates the extra degrees of freedom in the massive gravity get screened by non-linear self-interactions. The vDVZ discontinuity is just an artifact of the linear approximation. The second hurdle can be surpassed by carefully constructing models which avoid the BD ghosts, such as the Dvali-Gabadadze-Porrati (DGP) model \cite{Dvali:2000rv,Dvali:2000hr,Dvali:2000xg}, the new massive gravity \cite{Bergshoeff:2009hq}, and the de Rham-Gabadadze-Tolley (dRGT) gravity \cite{deRham:2010kj}.

Higher-spin fields $\varphi_{a_1 \cdots a_s}$ arise as generalisations of photons, gravitons and their massive counterparts. The construction of free massive higher-spin theory in flat space was first given by Singh and Hagen \cite{Singh:1974qz}. The massive fields are totally symmetric $\varphi_{a_1 \cdots a_s}=\varphi_{(a_1 \cdots a_s)}$, traceless $\varphi_{a_1 \cdots a_{s-2}ab}g^{ab} = 0$ (where $g^{ab}$ is the inverse metric), and they satisfy the transversality condition $\partial^a \varphi_{a a_2 \cdots a_s} = 0$. The free massless higher-spin theory was subsequently given by the Fronsdal programme \cite{Fronsdal:1978rb}, taking the massless limit of Singh-Hagen construction. The massless fields are instead double-traceless $\varphi_{a_1\cdots a_{s-4}a b c d}g^{ab}g^{cd} = 0$ and they exhibit gauge symmetries:
\begin{equation}
    \varphi_{a_1 \cdots a_s} \to \varphi_{a_1 \cdots a_s} + \partial_{(a_1} \lambda_{a_2 \cdots a_s)}
\end{equation} 
for some totally symmetric gauge parameter $\lambda_{a_1 \cdots a_{s-1}}$ which is traceless $\lambda_{a_1 \cdots a_{s-3}ab}g^{ab} = 0$. While free theories are valid, these higher-spin theories become problematic as soon as interactions are turned on. There are several no-go theorems which render the difficulties with massless spin $s > 2$ fields in flat background: the Weinberg low energy theorem \cite{Weinberg:1964ew} rules out massless higher-spin fields as long-range interaction carriers; the Coleman-Mandula theorem \cite{Coleman:1967ad} forbids the existence of non-trivial higher-spin conserved charges; the Weinberg-Witten theorem \cite{Weinberg:1980kq} and its generalisation \cite{Porrati:2008rm} prevent massless spin $s > 2$ fields from coupling minimally to the graviton in flat background. For massive higher-spin fields, interactions can lead to superluminal propagations in general (see, e.g.~\cite{Velo:1969txo,Velo:1972rt}). For general curved backgrounds, minimally coupled higher-spin fields do not have a well-posed initial value problem \cite{Wald:1984rg}. Also, if the curved background is not maximally symmetric, the minimal coupling of higher-spin fields fails because the commutator of covariant derivatives is proportional to the Riemann tensor (see e.g.~\cite{Christensen:1978md}). 

In order to circumvent all these no-go results, certain assumptions need to be abolished. For instance, one could allow non-minimal couplings, change the background from Minkowski to anti-de-Sitter (AdS) or de-Sitter (dS) spacetime, or go to three-dimensional spacetime. Another important lesson is that the consistency of massless higher-spin theories in dimensions $D \geq 4$ requires an infinite tower of fields with spin unbounded from above \cite{Fradkin:1986ka}. One important example of consistent higher-spin theory in 4D is the Vasiliev gravity \cite{Vasiliev:1990en,Vasiliev:1992av,Vasiliev:1995dn,Vasiliev:1999ba,Vasiliev:2003cph,Vasiliev:2004cp}, which is a non-linear theory of an infinite tower of interacting massless higher-spin fields formulated in AdS space.  It maintains gauge invariance, and offers insights into the AdS/CFT correspondence, where higher-spin theories in AdS space are dual to conformal field theories with conserved higher-spin currents \cite{Sezgin:2002rt,Klebanov:2002ja,Giombi:2009wh,Giombi:2011ya,Giombi:2012ms}. Another major example is 3D higher-spin gravity in terms of Chern-Simons theories \cite{Achucarro:1986uwr,Witten:1988hc} of gauge groups $\text{SL}(N,\mathbb{R})$, which allow finite species of higher-spin fields. These theories are topological, and the absence of local degrees of freedom in 3D further streamlines the dynamics and avoids inconsistencies found in higher-dimensional theories.  These 3D higher-spin theories admit black hole solutions \cite{Gutperle:2011kf,Kraus:2012uf,Castro:2016ehj}, which offer valuable clues in understanding the AdS$_3$/CFT$_2$ correspondence.

In this paper, we would like to understand what roles higher-spin fields play in the context of gravitational focusing. We take a bottom-up approach by further generalising the Raychaudhuri and focusing equations to include arbitrary spin $s \geq 2$ tensor fields $\varphi$ in the diffeomorphism-invariant Lagrangian $L$ and observe what problem would arise. For simplicity, we assume in our notation that only one higher-spin field is excited, however, the same procedure should apply to an infinite tower of higher-spin fields so long as the necessary sums converge.\footnote{It would be interesting to check when such convergence happens in a concrete model such as Vasiliev gravity.} The Lagrangian in consideration has the general form
\begin{equation}
    L = L\left(g ^{ab}, R _{abcd}, \nabla_{e_1} R_{abcd}, \cdots, \nabla_{(e_1 \cdots e_p)} R_{abcd}, \varphi_{a_1 \cdots a_s}, \nabla_{b_1} \varphi_{a_1 \cdots a_s}, \cdots, \nabla_{(b_1 \cdots b_q)} \varphi_{a_1 \cdots a_s}\right).
\end{equation}
For convenience, our discussion uses the terminology ``higher-spin'' to include massive gravitons. As in the case of $s \leq 1$, we again examine the focusing equation on Killing horizon backgrounds and switch on linear perturbations. We will also investigate the implications for the associated horizon entropy, which covers interesting cases such as dynamical black holes and cosmological horizons.

Before proceeding, we must note that the metric can transform non-trivially under higher-spin gauge transformations---this can alter causal structures and implies that causal horizons are no longer gauge-invariant \cite{Ammon:2011nk,Castro:2011fm}. Consequently, our current analysis is only relevant within the \emph{horizon gauge}, in which a Killing horizon is explicitly present. Throughout this paper, we will always assume a horizon gauge.

Unlike scalars and vectors, higher-spin fields introduce additional terms to the off-shell null-null component of the gravitational equation of motion $E_{vv}$ on the horizon:
\begin{equation}
 - 2 \pi E_{vv} \fheq \partial_v \Theta + \mathcal{L}_\xi \mathcal{P}_2, \label{eq:HS-Raychaudhuri}
\end{equation}
where, on the r.h.s., there is a desired $\partial_v \Theta$ term where the \emph{candidate} generalised expansion $\Theta$ still takes the form $\Theta = \partial_v \varsigma + D_i J^i$ in terms of an entropy density-current. In contrast to the $s\leq 1$ cases, there is an additional term $\mathcal{L}_\xi \mathcal{P}_2$, which is a collection of ``problematic'' terms which are not compatible with the structure of $\partial_v \Theta$ to be absorbed. They are linear in the perturbations of higher-spin field components, and they can be written in terms of a Lie derivative with respect to the Killing vector $\xi$, i.e., it vanishes on the stationary background. 

%To study the gravitational focusing, we source $E_{ab}$ with a stress-energy tensor $T_{ab}$ to obtain a candidate focusing equation:
%\begin{equation}
    %\partial_v \Theta = - 2 \pi T_{vv}  - \mathcal{L}_\xi \mathcal{P}_2. \label{eq:HS-focusing}
%\end{equation}

The main issue here is that $\mathcal{L}_\xi \mathcal{P}_2$ is not guaranteed to have a definite sign, and we call it an \emph{indefinite term}. Even if the NEC $E_{vv} = T_{vv} \geq 0$ is satisfied, the dynamically perturbed higher-spin field components could still defocus the generators on the horizon with respect to the candidate expansion $\Theta$ at the first order of perturbation! Moreover, the failure in proving a focusing theorem impedes the extraction of a non-decreasing Wall entropy because the structure of $T_{vv}$ is not integrable in the null direction.

In this paper, we propose a \emph{higher-spin focusing condition}, that is 
\begin{equation}
    \mathcal{L}_\xi \mathcal{P}_2 = 0.
\end{equation}

When this focusing condition is satisfied, the generalised expansion $\Theta$ is well-defined, and it obeys a focusing theorem $\partial_v \Theta \leq 0$ when NEC is provided. Moreover, the Wall entropy is well-defined as before, and the laws of thermodynamics continue to hold. The generalised expansion can still be interpreted as the divergence of the entropy density-current, and it is a consistent description of entropy production and redistribution.

It turns out that the existence of Wall entropy relies on a weaker condition---the \emph{averaged focusing condition}
\begin{equation}
    \mathcal{P}_2 = D_i \mathcal{J}^i
\end{equation}
i.e., the indefinite term can be non-zero locally, but it must be integrated to zero over a compact horizon. Subsequently, the generalised expansion is well-defined up to a spatial divergence, i.e., the light rays are allowed to anti-focus locally, but the total anti-focusing will cancel out over the entire compact horizon slice.

When this averaged condition is violated, it is hard even to discuss the concept of ``focusing'' because the Wall entropy is undefined, and the would-be generalised expansion has no physical meaning. One could object to this and insist that it is still a derivative of some ``entropy density-current''. But it is not associated to any entropy as it disregards any thermodynamic law. Hence, the ``generalised expansion'' is out of the scope of validity.

There is a potential way out when the averaged focusing condition is violated: the \emph{dynamical entropy} proposed by Hollands, Wald and Zhang \cite{Hollands:2024vbe}, which is an integral of the improved Noether charge of $\xi$ on the Killing horizon, is still well-defined and it satisfies both the first and the second laws even in the presence of higher-spin fields \cite{Visser:2024pwz}. But in this case, the dynamical entropy loses the interpretation as the Wall entropy for the associated generalised apparent horizon $\mathcal{A}$ \cite{HVWYZ}, simply because $\mathcal{A}$ cannot be defined.

The violation of the focusing condition in general higher-spin theories should be well-expected because most arbitrarily crafted higher-spin theories are pathological as discussed above. We thus speculate that this focusing condition could be related to a particular set of physical constraints on the higher-spin theories. As will be demonstrated later in the paper, we conjecture that \emph{physical higher-spin theories should possess sufficient symmetry to satisfy the focusing condition in a horizon gauge}. In other words, this condition should be necessary for the physical consistency of the theory.

\subsection{Plan of paper}
In Section \ref{sec:prelim}, the basic assumptions of this paper are illustrated, and the two major toolboxes---Gaussian null coordinates and covariant phase space formalism---are reviewed. The generalised focusing equation for spin $s \geq 2$ is derived, and three types of higher-spin focusing conditions are proposed in Section \ref{sec:GFE}. Section \ref{sec:entropy} demonstrates the implications of the focusing condition for two different horizon entropies: the Wall entropy and the dynamical entropy. Finally, in Section \ref{sec:discussion}, we speculate that focusing conditions could be necessary conditions for the physical consistency of higher-spin theories. Possible future directions are also discussed.

\section{Preliminaries}\label{sec:prelim}
We consider a $D$-dimensional spacetime with a bifurcate Killing horizon $\mathcal{H}$ consisting of a future horizon $\mathcal{H}^+$ and a past horizon $\mathcal{H}^-$ in diffeomorphism-invariant theories involving bosonic fields with arbitrary spin $s \in \mathbb{N}$. We in particular focus on the cases with $s \geq 2$.  In this paper, we limit our scope of discussion as follows:
\begin{enumerate}
    \item We define the ``nullness'' of any direction using the metric as usual;\footnote{Although, as mentioned in our previous work \cite{Wall:2024lbd}, we may need to consider Finsler geometry to determine the correct causal structure, we would leave this issue for future discussions.}
    \item We assume the existence of a solution with a regular bifurcate Killing horizon (we call it a \emph{stationary} black hole) with Killing vector $\xi$, and we normalise the surface gravity by $\kappa = 1$;
    \item We assume that the cross-sections of the Killing horizon are \emph{compact};
    \item We will always work in some \emph{horizon gauge} of the higher-spin fields where the Killing horizon is manifest;
    \item We only focus on \emph{first order non-stationary perturbations} (labeled by $\delta$) around \emph{stable} Killing horizon backgrounds, and we assume that $\delta \xi = 0$;
    \item We impose \emph{teleological} boundary condition such that all perturbations are switched off at future infinity so the horizon approaches stationarity;
    \item We require the matter field $\varphi$ to be \emph{smooth} on the horizon. The unperturbed matter field $\varphi$ satisfies the background Killing equation $\mathcal{L}_\xi \varphi = 0$. Fields with gauge symmetry are assumed to be in a gauge such that the Killing equation holds.\footnote{In general, the Killing equation holds up to a pure gauge transformation (see, e.g., \cite{Biswas:2022grc}). We will not consider such floating gauge degrees of freedom in this paper but will leave them to future work.}
\end{enumerate}

\subsection{Gaussian null coordinates}

Under the basic assumptions stated above, we can construct Gaussian null coordinates (GNC) around the future horizon $\mathcal{H}^+$ as follows. We choose $v$ to be the affine null parameter on $\mathcal{H}^+$, and the codimension-2 compact directions are labelled by $\{x^i\}$, $i=1,\cdots,D-2$. We extend off $\mathcal{H}^+$ by shooting ingoing affine null geodesics parametrised by $u$ and labelling $\mathcal{H}^+$ as $u=0$. (Notice that we follow the convention in \cite{Wall:2024lbd} where $\partial_u$ is past-directed.) In terms of coordinates $(u,v,x^i)$, the near-horizon metric reads 
\begin{equation}
    \dd{s^2} = 2 \dd{u} \dd{v} + u^2 F \dd{v^2} + 2 u \omega_i \dd{v} \dd{x^i} + h_{ij} \dd{x^i} \dd{x^j}
\end{equation}
where $F,\omega_i, h_{ij}$ are functions of $(u,v,x^i)$. We denote the codimension-2 $h_{ij}$-compatible covariant derivative as $D_i$. We will be working in GNC throughout this paper, and we assume $(u,v,x^i)$ are fixed under perturbations. 

\begin{figure}[H]
    \centering

    \tikzset{every picture/.style={line width=0.75pt}} %set default line width to 0.75pt        

    \begin{tikzpicture}[x=0.75pt,y=0.75pt,yscale=-1,xscale=1]
    
    %Curve Lines [id:da7469410763604614] 
    \draw [color={rgb, 255:red, 208; green, 2; blue, 27 }  ,draw opacity=1 ]   (299,65) .. controls (247.85,130.77) and (248.82,149.23) .. (299,215) ;
    \draw [shift={(261.28,134.31)}, rotate = 92.42] [color={rgb, 255:red, 208; green, 2; blue, 27 }  ,draw opacity=1 ][line width=0.75]    (8.74,-2.63) .. controls (5.56,-1.12) and (2.65,-0.24) .. (0,0) .. controls (2.65,0.24) and (5.56,1.12) .. (8.74,2.63)   ;
    %Straight Lines [id:da28612697242753526] 
    \draw [color={rgb, 255:red, 208; green, 2; blue, 27 }  ,draw opacity=1 ]   (170.41,183.59) -- (257.59,96.41) ;
    \draw [shift={(259,95)}, rotate = 135] [color={rgb, 255:red, 208; green, 2; blue, 27 }  ,draw opacity=1 ][line width=0.75]    (10.93,-3.29) .. controls (6.95,-1.4) and (3.31,-0.3) .. (0,0) .. controls (3.31,0.3) and (6.95,1.4) .. (10.93,3.29)   ;
    \draw [shift={(169,185)}, rotate = 315] [color={rgb, 255:red, 208; green, 2; blue, 27 }  ,draw opacity=1 ][line width=0.75]    (10.93,-3.29) .. controls (6.95,-1.4) and (3.31,-0.3) .. (0,0) .. controls (3.31,0.3) and (6.95,1.4) .. (10.93,3.29)   ;
    %Straight Lines [id:da49711972907857493] 
    \draw [color={rgb, 255:red, 208; green, 2; blue, 27 }  ,draw opacity=1 ]   (180.41,106.41) -- (247.59,173.59) ;
    \draw [shift={(247.59,173.59)}, rotate = 45] [color={rgb, 255:red, 208; green, 2; blue, 27 }  ,draw opacity=1 ][line width=0.75]    (10.93,-3.29) .. controls (6.95,-1.4) and (3.31,-0.3) .. (0,0) .. controls (3.31,0.3) and (6.95,1.4) .. (10.93,3.29)   ;
    \draw [shift={(180.41,106.41)}, rotate = 225] [color={rgb, 255:red, 208; green, 2; blue, 27 }  ,draw opacity=1 ][line width=0.75]    (10.93,-3.29) .. controls (6.95,-1.4) and (3.31,-0.3) .. (0,0) .. controls (3.31,0.3) and (6.95,1.4) .. (10.93,3.29)   ;
    %Straight Lines [id:da1234507755848453] 
    \draw [line width=1.5]    (119,235) -- (306.88,47.12) ;
    \draw [shift={(309,45)}, rotate = 135] [color={rgb, 255:red, 0; green, 0; blue, 0 }  ][line width=1.5]    (11.37,-3.42) .. controls (7.23,-1.45) and (3.44,-0.31) .. (0,0) .. controls (3.44,0.31) and (7.23,1.45) .. (11.37,3.42)   ;
    %Straight Lines [id:da8813428288683138] 
    \draw [line width=1.5]    (119,45) -- (306.88,232.88) ;
    \draw [shift={(309,235)}, rotate = 225] [color={rgb, 255:red, 0; green, 0; blue, 0 }  ][line width=1.5]    (11.37,-3.42) .. controls (7.23,-1.45) and (3.44,-0.31) .. (0,0) .. controls (3.44,0.31) and (7.23,1.45) .. (11.37,3.42)   ;
    %Straight Lines [id:da3869226710102489] 
    \draw [color={rgb, 255:red, 74; green, 144; blue, 226 }  ,draw opacity=1 ]   (214,140) ;
    \draw [shift={(214,140)}, rotate = 0] [color={rgb, 255:red, 74; green, 144; blue, 226 }  ,draw opacity=1 ][fill={rgb, 255:red, 74; green, 144; blue, 226 }  ,fill opacity=1 ][line width=0.75]      (0, 0) circle [x radius= 3.35, y radius= 3.35]   ;
    %Curve Lines [id:da4758785940969019] 
    \draw [color={rgb, 255:red, 208; green, 2; blue, 27 }  ,draw opacity=1 ]   (129,65) .. controls (180.52,131.13) and (180.52,149.6) .. (129,215) ;
    \draw [shift={(167.39,145.33)}, rotate = 272.1] [color={rgb, 255:red, 208; green, 2; blue, 27 }  ,draw opacity=1 ][line width=0.75]    (8.74,-2.63) .. controls (5.56,-1.12) and (2.65,-0.24) .. (0,0) .. controls (2.65,0.24) and (5.56,1.12) .. (8.74,2.63)   ;
    %Curve Lines [id:da27907733511506905] 
    \draw [color={rgb, 255:red, 208; green, 2; blue, 27 }  ,draw opacity=1 ]   (139,225) .. controls (205.13,174.21) and (223.6,174.58) .. (289,225) ;
    \draw [shift={(208.58,187.34)}, rotate = 357.03] [color={rgb, 255:red, 208; green, 2; blue, 27 }  ,draw opacity=1 ][line width=0.75]    (8.74,-2.63) .. controls (5.56,-1.12) and (2.65,-0.24) .. (0,0) .. controls (2.65,0.24) and (5.56,1.12) .. (8.74,2.63)   ;
    %Curve Lines [id:da5851765477210573] 
    \draw [color={rgb, 255:red, 208; green, 2; blue, 27 }  ,draw opacity=1 ]   (139,55) .. controls (205.13,106.52) and (223.6,106.52) .. (289,55) ;
    \draw [shift={(219.33,93.39)}, rotate = 177.9] [color={rgb, 255:red, 208; green, 2; blue, 27 }  ,draw opacity=1 ][line width=0.75]    (8.74,-2.63) .. controls (5.56,-1.12) and (2.65,-0.24) .. (0,0) .. controls (2.65,0.24) and (5.56,1.12) .. (8.74,2.63)   ;
    %Straight Lines [id:da2548043222724323] 
    \draw    (139,135) ;
    \draw [shift={(139,135)}, rotate = 0] [color={rgb, 255:red, 0; green, 0; blue, 0 }  ][fill={rgb, 255:red, 0; green, 0; blue, 0 }  ][line width=0.75]      (0, 0) circle [x radius= 3.35, y radius= 3.35]   ;
    %Straight Lines [id:da39093478809966675] 
    \draw [color={rgb, 255:red, 139; green, 87; blue, 42 }  ,draw opacity=1 ]   (232.5,122) ;
    \draw [shift={(232.5,122)}, rotate = 0] [color={rgb, 255:red, 139; green, 87; blue, 42 }  ,draw opacity=1 ][fill={rgb, 255:red, 139; green, 87; blue, 42 }  ,fill opacity=1 ][line width=0.75]      (0, 0) circle [x radius= 3.35, y radius= 3.35]   ;
    
    % Text Node
    \draw (319,40) node    {$v$};
    % Text Node
    \draw (320,232.5) node    {$u$};
    % Text Node
    \draw (248,93) node    {$\mathcal{H}^{+}$};
    % Text Node
    \draw (213.5,155.5) node  [color={rgb, 255:red, 74; green, 144; blue, 226 }  ,opacity=1 ]  {$\mathcal{B}$};
    % Text Node
    \draw (319,138.5) node  [color={rgb, 255:red, 208; green, 2; blue, 27 }  ,opacity=1 ]  {$\xi =v\partial _{v} -u\partial _{u}$};
    % Text Node
    \draw (114.5,135.5) node    {$\left\{x^{i}\right\}$};
    % Text Node
    \draw (241.5,138) node  [color={rgb, 255:red, 139; green, 87; blue, 42 }  ,opacity=1 ]  {$\mathcal{C}( v)$};
    % Text Node
    \draw (255.33,164.33) node    {$\mathcal{H}^{-}$};

    \end{tikzpicture}
    
\caption{Killing horizon in Gaussian null coordinates.}
\end{figure}

We label the bifurcation surface $\mathcal B$ by $u=v=0$. The GNC can be adapted \cite{Visser:2024pwz} so that the bifurcate Killing vector reads 
\begin{equation}
    \xi = v \partial_v - u \partial_u.
\end{equation}
Any tensor field $T$ is called \emph{stationary} iff it obeys the Killing equation $\mathcal{L}_\xi T = 0$, i.e., it is Lie transported by $\xi$.

The vector $\xi$ generates local boost transformation. We can then define the \emph{boost weight} \cite{Wall:2015raa} of a certain tensor \emph{component} in GNC by considering its transformation under a rigid boost $u \mapsto a u$, $v \mapsto v/a$. A component $T_{(w)}$ of weight $w$ transforms as $T_{(w)} \mapsto a^w T_{(w)}$. When all indices are lowered, the weight $w$ is equal to the number of $v$ indices minus the number of $u$ indices. We also call $w$ the number of ``net'' $v$-indices. In terms of the boost weight, the Lie derivative of $T_{(w)}$ with respect to $\xi$ reads
\begin{equation}
    (\mathcal{L}_\xi T)_{(w)} = (v \partial_v - u \partial_u + w) T_{(w)}. \label{eq:Lie-deriv}
\end{equation}

The benefits of boost weight are two-fold:
\begin{enumerate}[a.]
    \item It enables convenient accounting of first-order perturbations on $\mathcal{H}^+$: any positive-weight smooth tensor component vanishes on $\mathcal{H}^+$ at zeroth order, i.e., such components are of at least first order. Explicit products of positive-weight components are ignored on $\mathcal H^+$ because they are of at least second order. E.g., $E_{vv} \fheq \delta E_{vv} + \mathcal O(\delta^2)$, $A_{uvvi} B\indices{_v^i} \fheq \mathcal O(\delta^2)$, for some tensors $E_{ab}$, $A_{abcd}$, $B\indices{_a^b}$. See e.g., \cite{Wall:2024lbd} for a detailed discussion.
    \item It helps unravel the structure of dynamical positive-weight components on $\mathcal{H}^+$: in GNC, any ``net'' $v$-index in a weight-$w$ tensor component originates from either a $\partial_v$ or a $v$-index of the matter field component \cite{Wall:2024lbd}, if the tensor is constructed from the metric, the matter field and their derivatives.  E.g., $E_{vv} \supset \partial_v^2 \varsigma, \partial_v V_v, h_{vv}, \psi_{vuvvi}$, for some tensor $E_{ab}$ constructed out of the metric and the matter fields, some scalar $\varsigma$, and matter fields $V_a, h_{ab}, \psi_{abcde}$. For a detailed discussion on GNC decomposition, please refer to Claim 4 and Corollary 5 in \cite{Wall:2024lbd} based on results in \cite{Hollands:2022fkn}.
\end{enumerate}

\subsection{Covariant phase space equations}
Here, we derive some basic formulae in covariant phase space language in order to investigate the off-shell structure of the gravitational equations of motion. We take the standard treatment as in many previous works \cite{Lee:1990nz,Wald:1993nt,Iyer:1994ys,Wald:1999wa,Seifert:2006kv,Hollands:2012sf,Harlow:2019yfa}.

Start with the Lagrangian form $\mathbf L = L \bm \epsilon$ where $\bm \epsilon$ is the volume form. The Lagrangian density $L$ we are studying here is constructed from the most general coupling (scalar contraction) among the (inverse) metric $g^{ab}$, the Riemann tensor $R_{abcd}$ and its totally symmetric covariant derivatives, and some spin-$s$ ($s\in \mathbb{N}$) bosonic field $\varphi_{a_1 \cdots a_s}$ with its symmetric derivatives.\footnote{Any antisymmetric covariant derivative would contribute extra factors of Riemann tensor, using the Ricci identity.} Here, we do not need to impose symmetry or trace conditions on $\varphi_{a_1 \cdots a_s}$.\footnote{Such general treatment covers all the interesting cases: the symmetric traceless condition or totally antisymmetric condition can be imposed later to obtain the conventional higher-spin fields or $s$-form field. Also, this means the trace/symmetry conditions are not required to analyse the focusing equation.} The Lagrangian density $L$ reads 
\begin{widetext}
    \begin{equation}
        L = L\left(g ^{ab}, R _{abcd}, \nabla_{e_1} R_{abcd}, \cdots, \nabla_{(e_1 \cdots e_p)} R_{abcd}, \varphi_{A}, \nabla_{b_1} \varphi_{A}, \cdots, \nabla_{(b_1 \cdots b_q)} \varphi_{A}\right) \label{eq:lagrangian}
    \end{equation}
\end{widetext}
where, for conciseness of the notation, we have denoted $\nabla_{(e_1 \cdots e_p)} = \nabla_{(e_1} \cdots \nabla_{e_p)}$, and $A = a_1 \cdots a_s$ as a collection of $s$-indices for the matter field. For later convenience, we also denote $\phi \equiv (g, \varphi)$ collectively.

Varying $\mathbf L$: 
\begin{equation}
    \delta \mathbf L = \mathbf E \cdot \delta \phi + \dd{\mathbf \Theta[\phi, \delta \phi]}
\end{equation}
we obtain the equation of motion (EoM) forms:
\begin{equation}
    \mathbf E \cdot \delta \phi = \frac{1}{2} \mathbf E_{ab} \delta g^{ab} + \mathcal{\mathbf{E}}^{A} \delta \varphi_{A}
\end{equation}
where $\mathbf E_{ab} = E_{ab} \bm \epsilon$ and $\bm{\mathcal{E}}^{A} = \mathcal{E}^{A} \bm \epsilon$ are the EoM forms for $g$ and $\varphi$, respectively. The pre-symplectic potential $\mathbf \Theta$ is also obtained.\footnote{Do not confuse this with the generalised expansion $\Theta$ which is not bolded.} Note here we have adopted the convention that $E_{ab} = 2 (-g)^{-1/2} \delta \mathcal{I}/ \delta g^{ab}$ where $\mathcal{I} = \int \mathbf L$ is the action. 

In the case where we vary $\mathbf L$ through a diffeomorphism generated by a vector field $\zeta$, we replace $\delta$ with $\mathcal{L}_\zeta$ to obtain 
\begin{equation}
    \mathcal{L}_\zeta \mathbf L = \mathbf E \cdot \mathcal{L}_\zeta \phi + \dd{\mathbf \Theta[\phi, \mathcal{L}_\zeta \phi]}
\end{equation}
Subsequently, the \emph{off-shell $\zeta$-Noether current}\footnote{Off-shell means that we are not imposing the EoM $E=0$ and $\mathcal{E}=0$.} can be defined as 
\begin{equation}
    \mathbf J_\zeta = \mathbf \Theta_\zeta - \iota_\zeta \mathbf L
\end{equation}
where $\iota_\zeta$ is the contraction with $\zeta$ w.r.t.~first index. We then use the Cartan-Killing equation $\mathcal{L}_\zeta \mathbf L = \dd \iota_\zeta \mathbf L + \iota_\zeta \dd \mathbf L$, $\dd{\mathbf L} = 0$ for a top-form, and the expressions of $\mathcal{L}_\zeta g$ and $\mathcal{L}_\zeta \varphi$ in covariant derivatives to get 
\begin{equation}
    \dd{\mathbf J_\zeta} = \left(\left( E_{ab} - \chi_{ab} \right) \nabla^a \zeta^b - \mathcal{E}^{A} \zeta^b \nabla_b \varphi_{A} \right) \bm \epsilon
\end{equation}
where, for convenience, we have defined 
\begin{equation}
    \chi_{ab} \equiv \mathcal{E}\indices{_a^{a_2\cdots a_s}} \varphi_{b a_2 \cdots a_s} + \mathcal{E}\indices{^{a_1}_{a}^{a_3\cdots a_s}} \varphi_{a_1 b a_3 \cdots a_s} + \cdots + \mathcal{E}\indices{^{a_1\cdots a_{s-1}}_{a}} \varphi_{a_1\cdots a_{s-1}b}. \label{eq:matter-constraint}
\end{equation}
Note that $\chi_{ab}$ is not a symmetric tensor. Especially, for the case $s = 0$ (i.e., a scalar field), $\chi = 0$.

Invoking the \emph{generalised Bianchi identity}
\begin{equation}
    \nabla^a (E_{ab} - \chi_{ab}) + \mathcal{E}^{A} \nabla_b \varphi_{A} = 0,
\end{equation}
which is derived using the same methods as in \cite{Wall:2024lbd}, we obtain the off-shell conservation equation 
\begin{equation}
    \dd{(\mathbf J_\zeta + \mathbf C_\zeta)} = 0,
\end{equation}
where $\mathbf C_\zeta$ is the \emph{constraint form}, a $(D-1)$-form defined by
\begin{equation}
    \mathbf C_\zeta = (\chi^{ab} - E^{ab})\zeta_b \bm \epsilon_a\,. \label{eq:constraint-form}
\end{equation}
Here, $(\bm \epsilon_a)_{a_1 \cdots a_{D-1}}$ is a codimension-1 form obtained by fixing the first index $a$ in the volume form $\bm \epsilon_{a a_1 \cdots a_{D-1}}$. 

By the algebraic Poincar\'e lemma \cite{Wald:1990mme}, we define the \emph{off-shell $\zeta$-Noether charge} $\mathbf Q_{\zeta}$ such that 
\begin{equation}
    \mathbf J_\zeta + \mathbf C_\zeta = \dd{\mathbf Q_\zeta}.
\end{equation}

Varying this equation and assuming the diffeomorphism is field independent, i.e., $\delta \zeta = 0$,\footnote{At first order variation, $\delta \zeta$ will not contribute to the definition of black hole entropy, so we can set it to zero without loss of generality. This is studied in \cite{Visser:2024pwz} in detail.} we have the \emph{fundamental identity} that will be used in later sections:
\begin{equation}
    \delta \mathbf C_\zeta = \dd{(\delta \mathbf Q_\zeta - \iota_\zeta \mathbf \Theta[\phi, \delta \phi])} - \bm \omega[\phi; \delta \phi, \mathcal{L}_\zeta \phi] + \iota_\zeta \mathbf E \cdot \delta \phi\,\label{eq:magic-id}
\end{equation}
where $\bm \omega$ is the \emph{pre-symplectic form}:
\begin{equation}
    \bm \omega[\phi; \delta \phi, \mathcal{L}_\zeta \phi] = \delta \mathbf \Theta[\phi, \mathcal{L}_\zeta \phi] - \mathcal{L}_\zeta \mathbf \Theta[\phi, \delta \phi],
\end{equation} 
which vanishes for the Killing flow $\zeta = \xi$ because it is linear in $\mathcal{L}_\xi \phi = 0$.

\section{Generalised Focusing Equation}\label{sec:GFE}
\subsection{Gravitational focusing for spin $s \leq 1$}
To analyse the focusing of light rays on perturbed Killing horizons, we need to determine the structure of the null-null component of the gravitational equations of motion $E_{vv}$. It is a weight-2 tensor component that vanishes identically (even off-shell) on the background due to Killing symmetry. Hence, we only need to investigate the off-shell structure of the perturbed equation of motion $\delta E_{vv}$. For diffeomorphism-invariant theories with matter contents possessing spin $s \leq 1$, it is shown, using boost weight analysis and the fundamental identity, that \cite{Wall:2024lbd}
\begin{equation}
    - 2 \pi \delta E_{vv} \fheq \partial_v \left( \frac{1}{\sqrt{h}} \partial_v \delta \left( \sqrt{h}\, \varsigma \right) + D_i J^i \right) \equiv \partial_v \Theta \label{eq:low-spin-Raychaudhuri}
\end{equation}
in Gaussian null coordinates, where $\sqrt{h}$ is the area element of the codimension-2 horizon time slice, $\varsigma$ is the entropy density, and $J^i$ is the entropy current. In the last equality, we have identified the generalised expansion $\Theta$. It is the divergence of the entropy density-current vector $(\varsigma, J^i)$ on the horizon. This is the generalised linear Raychaudhuri equation. The gravitational focusing
\begin{equation}
    \partial_v \Theta \leq 0
\end{equation}
is ensured when $\delta E_{vv}$ is sourced by some external minimal matter field obeying NEC: $\delta E_{vv} = \delta T_{vv} \geq 0$. For GR, we have $\varsigma = 1$ and $J^i = 0$, so these reduce to the usual linear Raychaudhuri equation and the focusing of geometrical expansion.

\subsection{Generalised Raychaudhuri equation and indefinite terms for spin $s \geq 2$}

For spin $s \geq 2$ matter fields coupled to gravity, subtleties arise. Firstly, the $\delta \chi_{vv}$ in the constraint form \eqref{eq:constraint-form} no longer vanishes trivially off-shell as in the cases of $s \leq 1$. It should now be taken into account in the off-shell analysis. Secondly, $s \geq 2$ fields can have weight-2 or higher null polarisations which will contribute to terms in $\delta E_{vv}$ that cannot be written as a total $v$-derivative, i.e., they cannot be absorbed into the definition of generalised expansion. Furthermore, for a general theory, there is no obvious dynamical constraint on the sign of such terms. Hence, we call them \emph{indefinite terms}.

To obtain a generalised linear Raychaudhuri equation, we analyse the off-shell structure of the constraint form $\mathbf C_k$, which contains both $E_{vv}$ and $\chi_{vv}$. This is somewhat better than a direct decomposition of $E_{vv}$ because we can use the fundamental identity to prove certain properties for its structure. To unpack the constraint form, we work in GNC, where the boost weight argument \cite{Wall:2015raa} can be used to find the structure of $\delta (E_{vv} - \chi_{vv})$ on $\mathcal{H}^+$. Using Corollary 5 in \cite{Wall:2024lbd}, we obtain the preliminary form of the higher-spin generalised linear Raychaudhuri equation:
\begin{equation}
    - 2 \pi \delta (E_{vv} - \chi_{vv}) \fheq \partial_v \left( \frac{1}{\sqrt{h}} \partial_v \delta \left( \sqrt{h}\, \varsigma \right) + D_i \tilde J^i \right) + \mathcal{P}. \label{eq:Evv-chivv}
\end{equation}
Here, $\varsigma$ is the \emph{would-be entropy density}, $\tilde J^i$ is the manifest contribution to the \emph{would-be  entropy current},\footnote{The ``would-be''s suggest it is unclear whether we can extract a valid entropy formula from these when $\mathcal{P}$ is present.} and $\mathcal{P}$ is a collection of \emph{inauspicious terms} contributed by higher-spin fields, which have only one or zero total $v$-derivatives, and is not \emph{manifestly} a codimension-2 divergence. 

We now focus on the inauspicious terms because 1) they contain terms that cannot be absorbed into the generalised expansion and 2) the absorbed part may prevent the generalised expansion to be written in terms of an entropy density and an entropy current, which poses a problem for the extraction of horizon entropy. The detailed structure of $\mathcal{P}$ is the last element for fully understanding the generalised Raychaudhuri equation.

They can be divided into two types:
\begin{equation}
    \mathcal{P} = \partial_v \mathcal{P}_1 + \mathcal{P}_2 \label{eq:P}
\end{equation}
where there is one extractable $v$-derivative in the first term and none in the second. Terms like $\mathcal{P}_1$ already appeared in theories with vector fields \cite{Wall:2024lbd}, and $\mathcal{P}_2$ is only present for bosonic matter fields with spin $s \geq 2$. 

We know $\mathcal{P}_1$ has only one ``net'' $v$-index. We can expand this term as a sum
\begin{equation}
    \mathcal{P}_1 = \sum_{w=1}^{s}\sum_{I_w} X^{I_w}_{(1-w)}[\delta \tilde{\varphi}_{(w)}^{I_w}] \label{eq:P-1}
\end{equation}
where 
\begin{enumerate}
    \item $\delta \tilde{\varphi}^{I_w}_{(w)}$ labels a weight-$w$ ($1 \leq w \leq s$) component of the perturbed spin-$s$ field $\delta \varphi_A$ or its $u$-derivative (the decomposition follows again from Corollary 5 in \cite{Wall:2024lbd}), and the $I_w$ here labels the different species of the same weight. E.g., for a spin-4 field $\varphi_{abcd}$, weight-1 components include $\varphi_{vijk}$, $\varphi_{vuvi}$, $\partial_u \delta \varphi_{vijv}$, etc.
    \item $X^{I_w}_{(1-w)}$ is some theory-determined linear function involving codimension-2 spatial derivatives of $\delta \tilde{\varphi}_{(w)}^{I_w}$, which can be decomposed as 
    \begin{equation}
        X^{I_w}_{(1-w)}[\delta \tilde{\varphi}_{(w)}^{I_w}] = \sum_{r=0}^{n} \left( X^{I_w,r}_{(1-w)} \right)^{i_1\cdots i_r} D_{(i_1 \cdots i_r)}\left( \delta \tilde{\varphi}_{(w)}^{I_w} \right)
    \end{equation}
    where all the coefficients $X$ are weight-$(1-w)$ background quantities. Here, we kept the derivatives symmetric because we can use the codimension-2 spatial Ricci identity to absorb any anti-symmetric derivative as a codimension-2 Riemann tensor. Also, $n$ is the largest possible number of derivatives acting on $\delta \tilde{\varphi}$, which is $n = p + q + 2$ (see our Lagrangian \eqref{eq:lagrangian}).
\end{enumerate}

    Similarly, we can write, $\mathcal{P}_2$, which has two ``net'' $v$-indices as 
    \begin{equation}
        \mathcal{P}_2 = \sum_{w=2}^{s}\sum_{I_w} Y^{I_w}_{(2-w)}[\delta \tilde{\varphi}_{(w)}^{I_w}] \label{eq:P-2}
    \end{equation}
    where the $Y$'s are some weight-$(2-w)$ theory-determined linear functions, which are similar but different to $X$'s.

    Terms with $w = 1$ on $\delta \tilde{\varphi}$ will only appear in $\mathcal{P}_1$ by its weight structure, so we can single them out by writing 
    \begin{equation}
        \mathcal{P}_1 = \mathcal{P}_{1,1} + \mathcal{P}_{1,2}
    \end{equation}
    where 
    \begin{equation}
        \mathcal{P}_{1,1} = \sum_{I_1} X^{I_1}_{(0)}[\delta \tilde{\varphi}_{(1)}^{I_1}] \label{eq:P-1-1}
    \end{equation}
    and 
    \begin{equation}
        \mathcal{P}_{1,2} = \sum_{w=2}^{s}\sum_{I_w} X^{I_w}_{(1-w)}[\delta \tilde{\varphi}_{(w)}^{I_w}]. \label{eq:P-1-2}
    \end{equation}
    For vector fields, $\mathcal{P}_{1,1}$ is the only type of inauspicious terms present. When spin $s \geq 2$ fields are present, we notice immediately that $\mathcal{P}_{1,2}$ will mix with the terms in $\mathcal{P}_2$ because they share the same types of $\delta \tilde{\varphi}_{(w)}$.

    To determine the final structure of $\mathcal{P}_1$ and $\mathcal{P}_2$, we evaluate the identity \eqref{eq:magic-id} on the whole horizon $\mathcal{H}^+$ with $\zeta = \xi$ (the Killing vector) for a \emph{compactly supported} perturbation $\delta$\footnote{We can choose compact support because the identity is off-shell.} to get 
    \begin{equation}
        \int_{\mathcal{H}^+} \delta \mathbf C_\xi = \left( \int_{\mathcal{C}(\infty)} - \int_{\mathcal{C}(-\infty)} \right) \left( \delta \mathbf Q_\xi - \iota_\xi \mathbf \Theta [\phi, \delta \phi]\right) = 0,
    \end{equation}
    where the RHS vanishes due to the compact support of $\delta$. Unpacking the constraint form \eqref{eq:constraint-form}, we get  
    \begin{equation}
        \int_{v = -\infty}^{\infty} \dd{v} \int_{\mathcal{C}(v)} \dd[D-2]{x} \sqrt{h}\, v\, \delta (E_{vv} - \chi_{vv}) = 0.
    \end{equation}

    Plugging \eqref{eq:Evv-chivv} in, we observe that the terms containing $\varsigma$ and $\tilde J^i$ are integrated out respectively, one as a $v$-boundary term, and the other as a codimension-2 total derivative. We obtain the following equation for $\mathcal P_1$ and $\mathcal P_2$  
    \begin{equation}
        \int_{-\infty}^{+\infty} \dd{v} \int_{\mathcal{C}(v)} \dd[D-2]{x} \left( \mathcal{P}_1 - v \mathcal{P}_2 \right) = 0,
    \end{equation}
    which is valid for general compactly supported perturbation $\delta$. Substituting \eqref{eq:P-1} and \eqref{eq:P-2} in and integrate by parts in the $x^i$ directions for many times, we obtain 
    \begin{widetext}
    \begin{equation}
        \int_{-\infty}^{+\infty} \dd{v} \int_{\mathcal{C}(v)} \dd[D-2]{x} \sqrt{h} \left(\sum_{I_1} \delta \tilde{\varphi}_{(1)}^{I_1} \mathfrak{E}[X_{(0)}^{I_1}] + \sum_{w=2}^{s}\sum_{I_w} \delta \tilde{\varphi}_{(w)}^{I_w} \mathfrak{E}[X_{(1-w)}^{I_w} - v Y_{(2-w)}^{I_w}]\right) = 0 \label{eq:E-X-Y} 
    \end{equation}
    \end{widetext}
    where 
    \begin{equation}
        \mathfrak{E}[X_{(1-w)}^{I_w}] = \sum_{r=0}^{n} (-1)^r D_{(i_1 \cdots i_r)} \left( X_{(1-w)}^{I_w,r} \right)^{i_1 \cdots i_r}
    \end{equation}
    is the Euler-Lagrange equation for the function $X_{(1-w)}^{I_w}$ and similar for $Y$. Note that $\mathfrak{E}$ is linear in the coefficients. We conclude that 
    \begin{equation}
        \mathfrak{E}[X_{(0)}^{I_1}] = 0, \label{eq:E-X-0}
    \end{equation}
    and 
    \begin{equation}
        \mathfrak{E}[X_{(1-w)}^{I_w}] = v \mathfrak{E}[Y_{(2-w)}^{I_w}] \quad \text{for} \quad w \geq 2. \label{eq:E-X-v-Y}
    \end{equation}
    These hold because \eqref{eq:E-X-Y} should hold for any compactly supported perturbation, and different species of $\delta \tilde{\varphi}_{(w)}^{I_w}$ are independent. This is because, for general off-shell perturbations, we can independently specify the value of $\delta \varphi_A$ and their $u$-derivatives $\partial_u^k \delta \varphi_A$ on the horizon as there is no assumed constraint on $\varphi_A$ and on how we extend $\delta \varphi_A$ off the horizon. Also, the result should extend to non-compact perturbations (especially the on-shell ones), because these $\mathfrak{E}[X]$'s and $\mathfrak{E}[Y]$'s consist of background quantities, which do not depend on perturbations. 

    Plugging \eqref{eq:E-X-0} back into \eqref{eq:P-1-1} and rearranging using the product rule of $D_i$, we prove that $\mathcal{P}_{1,1}$ is in fact a codimension-2 divergence: 
    \begin{widetext}

    \begin{equation}
        \mathcal{P}_{1,1} = D_i J_{1,1}^i \quad \text{where} \quad J_{1,1}^i =  \sum_{I_1} \sum_{r=1}^{n} \sum_{r'=0}^{r-1}\left( (-1)^{r'} D_{i_1 \cdots i_{r'}} \left(X_{(0)}^{I_1,r}\right)^{(i i_1 \cdots i_{r-1})} D_{i_{r'+1} \cdots i_{r-1}} \delta \tilde{\varphi}_{(1)}^{I_1} \right).
    \end{equation}
    \end{widetext}
    This is precisely what we found in the case of vector field \cite{Wall:2024lbd}, where only $\mathcal{P}_{1,1}$ is present.

    The same treatment fails for $\mathcal{P}_{1,2}$ and $\mathcal{P}_2$, because the mixing between them would \emph{not} guarantee that the Euler-Lagrange equations $\mathfrak{E}[X_{(1-w)}^{I_w}]$ and $\mathfrak{E}[Y_{(2-w)}^{I_w}]$ vanish individually. Instead, this will give, for each $\delta \tilde{\varphi}_{(w)}^{I_w}$ with $w \geq 2$,
    \begin{equation}
        \begin{split}
            X_{(1-w)}^{I_w}[\delta \tilde{\varphi}_{(w)}^{I_w}] & = \delta \tilde{\varphi}_{(w)}^{I_w} \mathfrak{E}[X_{(1-w)}^{I_w}] + D_i \mathfrak{J}^i[X_{(1-w)}^{I_w}]\\
            & = v \delta \tilde{\varphi}_{(w)}^{I_w} \mathfrak{E}[Y_{(2-w)}^{I_w}] + D_i \mathfrak{J}^i[X_{(1-w)}^{I_w}]\\
            & = v Y_{(2-w)}^{I_w}[\delta \tilde{\varphi}_{(w)}^{I_w}] + D_i \mathfrak{J}_{I_w}^i
        \end{split}
    \end{equation}
    where $D_i \mathfrak{J}^i[X_{(1-w)}^{I_w}]$ are the codimension-2 total derivative obtained when calculating the Euler-Lagrange operator using integration by parts, and similar for that associated to $Y$, and we defined
    \begin{equation}
        \mathfrak{J}_{I_w}^i = \mathfrak{J}^i[X_{(1-w)}^{I_w}] - v \mathfrak{J}^i[Y_{(2-w)}^{I_w}]
    \end{equation}
    for convenience.
    
    When plugging back to the expression of the total inauspicious term \eqref{eq:P}, $\mathcal{P}_1$ has a $v$-derivative on it. So we consider
    \begin{equation}
        \partial_v \left( X_{(1-w)}^{I_w}[\delta \tilde{\varphi}_{(w)}^{I_w}] \right) = (v \partial_v + 1) Y_{(2-w)}^{I_w}[\delta \tilde{\varphi}_{(w)}^{I_w}] + \partial_v D_i \mathfrak{J}_{I_w}^i
    \end{equation}
    combining with each $Y_{(2-w)}^{I_w}[\delta \tilde{\varphi}_{(w)}^{I_w}]$ from $\mathcal{P}_2$, we have 
    \begin{equation}
        \begin{split}
            & \quad~ \partial_v \left( X_{(1-w)}^{I_w}[\delta \tilde{\varphi}_{(w)}^{I_w}] \right) + Y_{(2-w)}^{I_w}[\delta \tilde{\varphi}_{(w)}^{I_w}]\\
            & = (v \partial_v + 2) Y_{(2-w)}^{I_w}[\delta \tilde{\varphi}_{(w)}^{I_w}] + \partial_v D_i \mathfrak{J}_{I_w}^i\\
            & = \mathcal{L}_\xi \left( Y_{(2-w)}^{I_w}[\delta \tilde{\varphi}_{(w)}^{I_w}] \right) + \partial_v D_i \mathfrak{J}_{I_w}^i
        \end{split}
    \end{equation}
    where we have used the GNC expression \eqref{eq:Lie-deriv} of the Lie derivative of a weight-$2$ tensor component with respect to $\xi$, and the fact that $u=0$ on the horizon.

    Now, for the expression of the total inauspicious term \eqref{eq:P}, we have 
    \begin{equation}
            \mathcal{P} = \partial_v \mathcal{P}_1 + \mathcal{P}_2 = \partial_v D_i J^i_{\mathcal{P}} + \mathcal{L}_\xi \mathcal{P}_2
    \end{equation}
    where we used $\partial_v \sqrt{h} = 0$ on the background, and 
    \begin{equation}
        J^i_{\mathcal{P}} = J^i_{1,1} + \sum_{w = 2}^{s} \sum_{I_w} \mathfrak{J}_{I_w}^i.
    \end{equation}

    Finally, we obtain the higher-spin generalised Raychaudhuri equation on the horizon:
    \begin{equation}
            - 2 \pi \delta (E_{vv} - \chi_{vv}) \fheq \partial_v \left(\frac{1}{\sqrt{h}} \partial_v \delta \left(\sqrt{h}\, \varsigma \right) + D_i J^i \right) + \mathcal{L}_\xi \mathcal{P}_2 \label{eq:higher-spin-constraint}
    \end{equation}
    where $J^i = \tilde J^i + J^i_{\mathcal{P}}$ is the total entropy current. We call $\mathcal{L}_\xi \mathcal{P}_2$ the \emph{indefinite term}, which does not have a definite sign and prevents us from proving a focusing theorem. An example is given in the appendix \ref{app:toy-model} as an illustration of the existence of indefinite terms. 

\subsection{Higher-spin focusing condition}

We have obtained the generalised linear Raychaudhuri equation for diffeomorphism-invariant theories involving arbitrary bosonic fields. We have seen that problems arise when the spin of the matter field exceeds one. We rewrite the Raychaudhuri equation as
\begin{equation}
    \partial_v \Theta \fheq - 2 \pi \delta E_{vv} + 2 \pi \delta \chi_{vv} - \mathcal{L}_\xi \mathcal{P}_2, \label{eq:HS-Raychaudhuri-2}
\end{equation}
by identifying the \emph{candidate} generalised expansion as the divergence of the would-be entropy density-current
\begin{equation}
    \Theta = \frac{1}{\sqrt{h}} \partial_v \delta \left(\sqrt{h}\, \varsigma \right) + D_i J^i.
\end{equation}
In the equation, we see two worrisome terms. The first one, $\delta \chi_{vv}$, is not too bad. It can be ruled out by concentrating on the NEC-compliant \emph{external} minimal perturbations that only source the gravitational field:\footnote{One could turn on an external scalar sector $\frac{1}{2} g^{ab} \nabla_a f \nabla_b f$ as a part of the perturbation. This would only contribute to the gravitational equation of motion, keeping $\chi_{ab} = \delta \chi_{ab} = 0$ at the first order.}
\begin{equation}
    \delta E_{vv} = \delta T_{vv} \geq 0, \qquad \delta \mathcal{E}^A = 0.
\end{equation}
The second one is worse because it is not constrained in general, so it cannot have a definite sign. This suggests there is no focusing theorem if we pick an arbitrary theory of gravity and higher-spin fields. Furthermore, the second law of horizon thermodynamics would fail as a negative generalised expansion is not forbidden, and all kinds of exotic consequences could occur. We will refer to this as the \emph{gravitational focusing problem} for higher-spin theories.

There are two different interpretations of the gravitational focusing problem for general higher-spin theories:
\begin{enumerate}
    \item We accept that this $\Theta$ is the generalised expansion, and we find that $\partial_v \Theta$ has indefinite sign;
    \item We do not accept this $\Theta$ and assert that a generalised expansion is undefined because eq.~\eqref{eq:HS-Raychaudhuri-2} does not have the right form for a linear focusing equation due to the indefinite term.
\end{enumerate} 
As we will see in the next section, the discussion of horizon entropy will pick the latter as the preferred interpretation.

We now propose a \emph{focusing condition}, which could also be thought of as constraints on the higher-spin fields/theories. If we take the first approach, we would require
\begin{equation}
    \mathcal{L}_\xi \mathcal{P}_2 \geq 0. 
\end{equation}
However, one can spot an immediate problem: $\mathcal{L}_\xi \mathcal{P}_2$ is linear in the perturbed components of higher-spin fields $\delta \tilde{\varphi}_{(w)}^{I_w}$, and we can always reverse the sign of the perturbation parameter $\varepsilon \to - \varepsilon$: we can have $\varphi \to \varphi - \varepsilon \delta \varphi$ instead of $\varphi \to \varphi + \varepsilon \delta \varphi$. Therefore, any $\mathcal{L}_\xi \mathcal{P}_2 \geq 0$ would become $\mathcal{L}_\xi \mathcal{P}_2 \leq 0$ and the focusing theorem fails again. We are then left with the true focusing condition
\begin{equation}
    \mathcal{L}_\xi \mathcal{P}_2 = 0
\end{equation}
which is consistent with the second interpretation: the generalised expansion only exists when the indefinite term vanishes. Because $\mathcal{P}_2$ is a linear function of $\delta \tilde{\varphi}_{(w)}^{I_w}$ and their spatial derivatives with background-valued coefficients, we see that the Lie derivative only acts on the perturbations: 
\begin{equation}
    \mathcal{L}_\xi \mathcal{P}_2 = \mathcal{L}_\xi \left( \sum_{w=2}^{s}\sum_{I_w} Y_{(2-w)}^{I_w}[\delta \tilde{\varphi}_{(w)}^{I_w}] \right) =  \sum_{w=2}^{s}\sum_{I_w} Y_{(2-w)}^{I_w}[\mathcal{L}_\xi \delta \tilde{\varphi}_{(w)}^{I_w}].
\end{equation}
For the stationary background, Killing symmetry ensures the vanishing of the indefinite term, and it admits a constant generalised expansion that is consistent with the assumptions.

The focusing condition requires a linear combination of the perturbations of field components and their spatial derivatives to vanish. Therefore, we can convert this into conditions on the coefficients $Y_{(2-w)}^{I_w}$ and the components $\delta \tilde \varphi_{(w)}^{I_w}$. One way to achieve the focusing condition is to require the field components $\tilde \varphi_{(w)}^{I_w}$ to be zero on the horizon using gauge redundancies of the higher-spin fields. Such a gauge condition should be protected under perturbations. Another possibility is that the higher-spin theories should exhibit specific symmetries to make the field components linearly dependent (e.g., through traceless conditions). With a special structure, cancellations among the coefficients could be possible. Alternatively, we can demand that the higher-spin theories have peculiar structures which ensure all the coefficients in $Y_{(2-w)}^{I_w}$ to be zero. There is yet another route to make $\mathcal{L}_\xi \mathcal{P}_2$ vanish: we can impose a stationarity condition on the perturbations to weight $w \geq 2$ field components: $\mathcal{L}_\xi \delta \tilde \varphi_{(w)}^{I_w} = 0$. However, this requirement may be problematic, as it imposes the Killing equation on only specific components of a tensor, rendering the condition non-covariant. We will not consider this option further.

In summary, neglecting the non-covariant choice, there are two types of higher-spin focusing conditions that could solve the gravitational focusing problem:
\begin{enumerate}[leftmargin=\leftmargin+\parindent,align=left]
    \item[Type G.] $\delta \tilde \varphi_{(w)}^{I_w} = 0$ for $w \geq 2$ on the horizon: This is a \emph{gauge condition} that requires weight $w \geq 2$ components $\tilde \varphi_{(w)}^{I_w}$ to be gauged away in some horizon gauge.
    \item[Type S.] $\mathcal{P}_2 = 0$: This is a \emph{structural condition} that requires the higher-spin field and theory to have specific structures and/or symmetries. This covers the subcase $Y_{(2-w)}^{I_w} = 0$. 
\end{enumerate}

Type S focusing condition is a stringent constraint---the higher-spin theories may need very specific types of interactions to satisfy this condition. At this stage, the coefficients in $Y_{(2-w)}^{I_w}$ could only be determined by brute-force calculation once a Lagrangian is given. Therefore, we will postpone discussing this condition in our future work. Type G condition is much more universal: it is a generic expectation that massless higher-spin fields would have lots of gauge redundancies, and it is speculated that one can choose a gauge (among the horizon gauges) in which the weight-2 or higher components could be set to zero. Any perturbation should preserve the gauge, as there should not be any discontinuous jump in the number of physical degrees of freedom upon perturbation. For massive higher-spin fields, it might be possible to carry out a consistent St\"uckelberg treatment (see, e.g.~\cite{deRham:2010kj,Porrati:2008ha}) which preserves the correct number of propagating degrees of freedom (i.e., no ghost modes are introduced). In this procedure, a massive higher-spin field can be decomposed into a tower of massless fields with lower spins, and the gravitational focusing problem of massive fields can be converted into that of massless fields.

\section{Horizon entropy and higher-spin focusing condition}\label{sec:entropy}
\subsection{Focusing problem and validity of Wall entropy}
Wall entropy is defined by integrating the null-null component of the gravitational equation of motion on a compact time-slice $\mathcal{C}$ of the horizon \cite{Wall:2015raa}: 
\begin{equation}
    \partial_v^2 S_\text{Wall} = - 2 \pi \int_\mathcal{C} \dd[D-2]{x} \sqrt{h}\, E_{vv}. \label{eq:Wall-def}
\end{equation}
The validity of such an indirect definition through a second derivative highly relies on the \emph{integrability} of $E_{vv}$ in $v$, i.e., whether the integral of $E_{vv}$ has two manifest $\partial_v$'s. For gravity coupled to spin $s \leq 1$ bosonic matter fields, the structure of $E_{vv}$ is integrable---it admits a generalised expansion
\begin{equation}
    \Theta = \frac{1}{\sqrt{h}} \partial_v \left( \sqrt{h}\, \varsigma \right) + D_i J^i
\end{equation}
(see eq.~\eqref{eq:low-spin-Raychaudhuri}) and the entropy can be calculated as 
\begin{equation}
    S_{\text{Wall}} = \int_\mathcal{C} \dd[D-2]{x} \sqrt{h}\, \varsigma
\end{equation}
where the divergence of entropy current $D_i J^i$ is integrated out on the compact slice $\mathcal{C}$. The Wall entropy enjoys the following properties:
\begin{enumerate}
    \item \emph{Locality}: it is a codimension-2 integral of local geometrical quantities on an arbitrary time-slice $\mathcal{C}$ of the horizon, and it only depends on the data at one instant of null time $v$;
    \item \emph{Non-stationarity and uniqueness}: it contains non-stationary corrections to the Wald entropy and is invariant under JKM ambiguities \cite{Wall:2015raa,Wall:2024lbd,Kar:2024dqk};
    \item \emph{First law}: it obeys the physical version of first law when integrated on the future horizon:
    \begin{equation}
        \frac{\kappa}{2 \pi} \Delta S_{\text{Wall}} = \int_{0}^{\infty} \dd{v} \int \dd[D-2]{x} \sqrt{h}\, T_{ab} k^a \xi^b = \Delta M - \Omega_{\mathcal{H}} \Delta J;
    \end{equation}
    \item \emph{Second law}: it satisfies a linearised second law when $E_{vv}$ is sourced by a NEC-compliant energy density $T_{vv} \geq 0$:
    \begin{equation}
        \partial_v S_{\text{Wall}} \geq 0
    \end{equation}
    given the teleological boundary condition $\partial_v S_\text{Wall} \to 0$ at infinity;
    \item \emph{GNC gauge invariance}: though the Wall entropy is extracted from a specific gauge choice of GNC, it is proven to be gauge invariant under coordinate transformations to different GNCs at the linear order of perturbations\footnote{Wall entropy is only defined to this order anyway.} \cite{Hollands:2022fkn,Bhattacharyya:2022njk}.
\end{enumerate}
Ultimately, these properties directly result from the gravitational focusing on the horizon.

When generalising the Wall entropy to diffeomorphism-invariant theories involving spin $s \geq 2$ matter fields, we demonstrate that the higher-spin focusing problem poses a significant obstacle. For the following discussion, we assume the external perturbations do not source the higher-spin equations of motion, so $\delta \chi_{ab} = 0$.\footnote{For $\delta \chi_{ab} \neq 0$, it would contribute an extra $\Phi \delta Q$-like term in the first law, but this may only make sense when the matter field in consideration is a $p$-form gauge field. To allow a non-vanishing charge term, the gauge field has to be divergent at the bifurcation surface to ensure a non-zero electric potential $\Phi$. However, the current discussion of Wall entropy is based on the assumption that all the dynamical fields are smooth over the horizon. We will leave such cases to future work.} 

To find the Wall entropy from its defining relation eq.~\eqref{eq:Wall-def}, we invert $\partial_v^2$ using its Green's function to obtain the Wall entropy at a horizon time-slice $\mathcal{C}(v)$:
\begin{equation}
    S_\text{Wall}(v) = S_\text{Wall}(\infty) - 2 \pi \int_{v}^{\infty} \dd{v'} \int_{\mathcal{C}(v')} \dd[D-2]{x} \sqrt{h}\, (v' - v) E_{vv}.
\end{equation}
This is a non-local expression because it expresses the entropy at $\mathcal{C}(v)$ in terms of data to the future of $\mathcal{C}(v)$ all the way up to $\mathcal{C}(\infty)$. Only integrable $E_{vv}$ (e.g. for spin $s\leq 1$) can yield a local $S_\text{Wall}(v)$ that depends on data at $\mathcal{C}(v)$ only. Plugging in the generalised linear Raychaudhuri equation
\begin{equation}
   - 2\pi E_{vv} = \partial_v \Theta + \mathcal{L}_\xi \mathcal{P}_2
\end{equation}
we have
\begin{equation}
    S_\text{Wall}(v) = S_\text{Wall}(\infty) + \left[ \int_{\mathcal{C}(v')} \dd[D-2]{x}\sqrt{h}\, \varsigma \right]_{v'=\infty}^v + v \int_{\infty}^{v} \dd{v'}\int_{\mathcal{C}(v')} \dd[D-2]{x} \sqrt{h}\, \mathcal{P}_2
\end{equation}
where we have used integration by parts and the teleological boundary condition that $\partial_v \varsigma, \mathcal{P}_2 \to 0$ as $v \to \infty$. The gravitational focusing problem for spin $s \geq 2$ manifests itself in terms of a non-local entropy formula on the horizon, as a result of $\mathcal{P}_2 \neq 0$. In fact, at the level of entropy, the focusing problem can be reduced to a weaker form. The local focusing condition can be violated, but we require an \emph{averaged focusing condition} as the condition for integrability: the existence of Wall entropy is guaranteed when
\begin{equation}
    \mathcal{P}_2 = D_i \mathcal{J}^i
\end{equation} 
which can be integrated out on a compact horizon time-slice. This condition is quite stringent because it requires the higher-spin theories to have a very specific structure.

When the averaged focusing condition is violated, we could make another attempt to circumvent the focusing problem: we modify the definition of Wall entropy to be 
\begin{equation}
    \tilde S_{\text{Wall}} = \int_{\mathcal{C}} \dd[D-2]{x} \sqrt{h}\, \varsigma.
\end{equation}
However, the problem persists because this modified entropy does not obey the second law due to anti-focusing. We have traded the locality with the second law, but losing either is nonsensical. Therefore, our conclusion echoes the second interpretation of the focusing problem in the previous section: when the averaged focusing condition is violated, the Wall entropy is undefined. The ``generalised expansion'' $\Theta$ is subsequently ill-defined because the associated ``entropy density-current'' does not respect any thermodynamic law, and it does not contain any physical information about the null geodesic congruence.

When an averaged but local focusing is achieved, the Wall entropy satisfying the five properties is defined, but the generalised expansion is only defined up to a spatial divergence. In this case, the light rays are allowed to anti-focus locally, but the total anti-focusing will cancel out over the entire compact horizon slice.

\subsection{A potential way out: dynamical entropy}

In order to circumvent the focusing condition altogether, we can use the \emph{dynamical black hole entropy} defined recently by Hollands, Wald and Zhang \cite{Hollands:2024vbe}. It is defined as 
\begin{equation}
    S_\text{dyn} = 2 \pi \int_{\mathcal{C}(v)} \left( \mathbf Q_\xi - \iota_\xi \mathbf B \right)
\end{equation}
where $\mathbf Q_\xi$ is the boost Noether charge, and $\mathbf B$ is defined by extracting $\delta$ from the pre-symplectic potential: $\mathbf \Theta[\delta] \fheq \delta \mathbf B$, which is proven in \cite{Hollands:2024vbe} for pure gravity. In \cite{Visser:2024pwz}, we have generalised this proof and, hence, the existence of dynamical entropy to all bosonic matter fields. This suggests that the dynamical entropy is, by definition, a codimension-2 integral that is local in the null direction. To find an expression of the dynamical entropy in terms of the would-be Wall entropy density $\varsigma$ and the indefinite term $\mathcal{P}_2$, we use the expression of its first derivative at the first order of perturbation:
\begin{equation}
    \begin{split}
        \partial_v S_\text{dyn} & = 2 \pi \int_{\mathcal{C}(v)} \dd[D-2]{x} \sqrt{h}\, v\, (E_{vv} - \chi_{vv}) = \partial_v \int_{\mathcal{C}(v)} \dd[D-2]{x} \sqrt{h} \left( \varsigma - v \partial_v \varsigma - v^2 \mathcal{P}_2 \right)
    \end{split}
\end{equation}
where we have plugged in the generalised Raychaudhuri equation \eqref{eq:higher-spin-constraint} and used 
\begin{equation}
    v \mathcal{L}_\xi \mathcal{P}_2 = v^2 \partial_v \mathcal{P}_2 + 2 v \mathcal{P}_2 = \partial_v (v^2 \mathcal{P}_2).
\end{equation}

The dynamical entropy can then be derived up to the first order as 
\begin{equation}
    S_\text{dyn} = \int_{\mathcal{C}(v)} \dd[D-2]{x}\sqrt{h}\, \left( \varsigma - v \partial_v \varsigma - v^2 \mathcal{P}_2 \right).
\end{equation}
This is a codimension-2 integral local in $v$ regardless of the focusing condition. 

The Noether charge picture of the dynamical entropy guarantees its first law \cite{Hollands:2024vbe,Visser:2024pwz}. It also obeys the second law because 
\begin{equation}
    \partial_v S_\text{dyn} = 2 \pi \int_{\mathcal{C}(v)} \dd[D-2]{x} \sqrt{h}\, v\, T_{vv} \geq 0
\end{equation}
where we used the perturbed equations of motion and the NEC. 

When the averaged focusing condition is satisfied, we recover the neat relation \cite{Hollands:2024vbe}
\begin{equation}
    S_\text{dyn} = (1 - v \partial_v) S_\text{Wall}.
\end{equation}
In GR, this relation has a profound implication: the Wall entropy reduces to the Bekenstein-Hawking entropy and at the first order of perturbation
\begin{equation}
    S_\text{dyn}[\mathcal{C}] = (1 - v \partial_v) S_{\text{BH}}[\mathcal{C}] = S_{\text{BH}}[\mathcal{A}]
\end{equation}
where $\mathcal{C}$ is a time-slice of the Killing horizon, while $\mathcal{A}$ is the associated \emph{apparent horizon} that lives in the interior of $\mathcal{C}$ with an affine distance of linear order in the perturbation parameter. In plain language, the dynamical entropy of the event horizon is the area (over four) of the apparent horizon! It locates the local boundary of the dynamical horizon (e.g., the boundary of a dynamical black hole). 

For general diffeomorphism-invariant theory, we speculate that a similar relation would hold, given that a \emph{generalised apparent horizon} $\mathcal{A}$ can be defined to linear order \cite{HVWYZ} (see also \cite{Kong:2024sqc}):
\begin{equation}
    S_\text{dyn}[\mathcal{C}] \overset{?}{=} S_{\text{Wall}}[\mathcal{A}].
\end{equation}
We can see that the focusing condition is crucial for the existence of a generalised expansion $\Theta$, hence a $\Theta = 0$ surface. In other words, although dynamical entropy can circumvent the focusing condition, it loses the interpretation of entropy for the apparent horizon when the focusing condition is not satisfied. We will leave the detailed analysis of this, especially the role of $\mathcal{P}_2$ in $S_\text{dyn}$, to future work.

\section{Discussion}\label{sec:discussion}
\subsection{Speculations: focusing as constraints}
    In this paper, we have pursued whether there is a focusing theorem for general diffeomorphism-invariant theories involving higher-spin fields. Answering this question has led to a mutual test among the higher-spin fields, gravitational focusing, and the principles of horizon thermodynamics. We have seen that to ensure the focusing of light and an increasing Wall entropy, the higher-spin fields/theories are constrained by the focusing condition. With the belief that a focusing theorem should hold for physical theories of gravity, and that the Wall entropy is a valid and direct generalisation of Bekenstein-Hawking entropy,\footnote{This is supported by its validity for pure gravity, scalar fields and vector fields, and its agreement with the holographic entanglement entropy (Dong entropy) \cite{Dong:2013qoa} in $f(\text{Riemann})$ theory.} we speculate that physically consistent higher-spin theory should obey the focusing condition, at least on a bifurcate Killing horizon background where we have managed to define the notion of ``focusing''. For a universal consideration, we focus on the Type G condition, which does not require detailed knowledge of interactions in a theory.
    
    Massless fields have a high degree of gauge symmetry. We expect them to have enough redundancies to gauge away weight $w \geq 2$ field components on the horizon, and this gauge should be preserved at least for linear dynamical perturbations so that the Type G condition is satisfied:
    \begin{equation}
        \delta \tilde{\varphi}^{I_w}_{w \geq 2} \fheq 0.
    \end{equation}
    In other words, we expect there exists a horizon gauge where the Type G focusing condition is met.

    Notice that these are requirements on the horizon. If there is enough symmetry to set some field component to zero \emph{in a neighbourhood} of the horizon (or globally, if possible), then its $u$-derivatives on the horizon are guaranteed to be zero. Therefore, we may instead ask the field components to satisfy
    \begin{equation}
        \delta \varphi_{w\geq 2}^{I_w} = 0
    \end{equation}
    in a neighbourhood of $\mathcal{H}^+$. 

    ~

    \noindent {\textbf{Massless graviton and $p$-form fields}}

    \vspace{0.25em}

    Massless gravitons and $p$-form fields trivially satisfy the focusing condition no matter what interactions they have. 

    A graviton $\gamma_{ab}$, which can also be seen as a perturbation to the metric, should satisfy the GNC conditions imposed on the metric, 
    \begin{equation}
        \gamma_{vv} = u^2 f(u,v,x) \fheq 0,
    \end{equation}
    i.e., the horizon remains null with respect to it. The focusing condition is automatically satisfied. This also provides a trivial self-consistency check that the previous discussions of gravitational focusing and Wall entropy are valid for pure gravity.

    For a $p$-from field $B_{[a_1\cdots a_p]}$, there could only be at most one $v$-index on any component by antisymmetry, so $\mathcal{P}_2$ vanishes identically. The focusing condition is satisfied because $p$-form fields are effectively spin-1, the same as Proca fields discussed in \cite{Wall:2024lbd}.
    
    ~

    \noindent \textbf{Symmetric higher-spin fields} 

    \vspace{0.25em}
    
    For symmetric spin-$s$ fields, the focusing condition can be satisfied by imposing the following gauge conditions: 
    
    Spin-2: $\delta \varphi_{vv} = 0$;
        
    Spin-3: $\delta \varphi_{vvv} = \delta \varphi_{vvi} = 0$;

    Spin-4: $\delta \varphi_{vvvv} = \delta \varphi_{vvvi} = \delta \varphi_{vvij} = \delta \varphi_{vvvu} = 0$;

    And so on. 

    Intuitively, these conditions should be physically acceptable, at least for massless fields, because we expect them to have only non-zero transverse polarisations (at least in a particular gauge). In Appendix \ref{app:3d}, we give an example: a 3D higher-spin black hole with spin-3 field \cite{Gutperle:2011kf}. It is shown that in the black hole gauge, the spin-3 field satisfies the focusing condition on the horizon.
    
    For massive higher-spin fields, it seems complicated for them to directly satisfy the Type G condition. Nevertheless, as long as the theory allows a consistent St\"uckelberg process, we can always reduce the problem to that of a tower of massless spin $s\geq 2$ fields with gauge symmetry plus a vector field. We can then resort to transversality conditions for massless fields and use our previous results in \cite{Wall:2024lbd} for the vector fields.

    For example, in a flat background (where we could consider Rindler horizons), the massive spin-2 field with mass $m$ can be redefined as 
    \begin{equation}
        \varphi_{ab} \to  \Phi_{ab} = \varphi_{ab} + \frac{1}{m} \partial_{(a} V_{b)}
    \end{equation}
    where $V_a$ is the St\"uckelberg vector field. They have gauge transformations
    \begin{equation}
        \begin{split}
            \varphi_{ab} & \to \varphi_{ab} + \partial_{(a} \lambda_{b)},\\
            V_{a} & \to V_{a} - m \lambda_{a}.
        \end{split}
    \end{equation}
    We can then pick a gauge via 
    \begin{equation}
        \partial_v \lambda_v = - \varphi_{vv}
    \end{equation}
    to make the redefined $\varphi_{ab}$ to satisfy the focusing condition. The vector field $\lambda_a$ will not cause any difficulty in the focusing equation. For general curved backgrounds, a non-linear realisation of the St\"uckelberg trick for massive gravitons is provided in the dRGT gravity \cite{deRham:2010kj}. Hence, it should be possible to gauge the dRGT massive graviton so that it satisfies the focusing condition.
    
    For a spin-$s$ massive field with mass $m$, at least around the flat background, we can unpackage it into a tower of massless St\"uckelberg fields (see, e.g.~\cite{Porrati:2008ha})
    \begin{equation}
        \varphi_{a_1 \cdots a_s} \to (\varphi^{(s)}_{a_1 \cdots a_{s}}, \varphi^{(s-1)}_{a_1 \cdots a_{s-1}}, \cdots, \varphi^{(2)}_{ab}, V_a),
    \end{equation}
    which have the following gauge transformation rules:
    \begin{equation}
        \begin{split}
            \varphi^{(s)}_{a_1 \cdots a_{s}} & \to \varphi^{(s)}_{a_1 \cdots a_{s}} + \partial_{(a_1} \lambda^{(s-1)}_{a_2 \cdots a_s)}\\
            \varphi^{(s-1)}_{a_1 \cdots a_{s-1}} & \to \varphi^{(s-1)}_{a_1 \cdots a_{s-1}} - m \lambda^{(s-1)}_{a_1 \cdots a_{s-1}} + \partial_{(a_1} \lambda^{(s-2)}_{a_2 \cdots a_{s-1})}\\
            & \cdots\\
            \varphi^{(2)}_{ab} & \to \varphi^{(2)}_{ab} - (s-2)m \lambda^{(2)}_{ab} + \partial_{(a} \lambda^{(1)}_{b)}\\
            V_{a} & \to V_a - (s-1)m \lambda^{(1)}_a
        \end{split}
    \end{equation}
    where $\lambda^{(s-k)}_{a_1 \cdots a_{s-k}}$ ($k = 1, \dots, s-1$) are the gauge parameters. We can tune these parameters to make the unpacked massless higher-spin fields $\varphi^{(2)}, \dots \varphi^{(s)}$ satisfy the focusing condition. The vector field left can be treated using previous results in \cite{Wall:2024lbd}. Using the St\"uckelberg trick, the focusing condition comes with the price that the gauge parameters can show up in the non-zero field components of $\varphi^{(2)}, \dots, \varphi^{(s)}$, which could, in turn, appear in the horizon entropy density extracted from the focusing equation. 

    ~

   \noindent \textbf{Counting degrees of freedom} 

   \vspace{0.25em}
   
   To further support our speculation, we will quickly count the degrees of freedom (DoF) to show that the number of conditions required is smaller than that of redundancies in higher-spin fields. This may offer insights into general curved backgrounds where we do not know whether we can resort to the transversality condition and/or a consistent St\"uckelberg trick.
    
    The focusing condition require $n(D,s)$ global conditions for a spin-$s$ totally symmetric tensor field in $D$ dimensions, where
    \begin{equation}
        n(D,s) = \sum_{k_u = 0}^{\lfloor s/2\rfloor - 1} \sum_{k_v = k_u + 2}^{s-k_u}\binom{D - 2 + s - k_v - k_u - 1}{s - k_v - k_u}.
    \end{equation}

    We take 4D higher-spin fields (e.g.~Fronsdal higher-spin fields \cite{Fronsdal:1978rb}) as an example. The number of constraints evaluates to 
    \begin{equation}
        n(4,s) = \frac{1}{6} \left\lfloor \frac{s}{2}\right\rfloor  \left(\left\lfloor \frac{s}{2}\right\rfloor  \left(4 \left\lfloor \frac{s}{2}\right\rfloor -6 s-3\right)+3 s (s+1)-1\right).
    \end{equation}
    Assuming the number of physical DoF is fixed when we move from the flat spacetime to curved ones (as we do not want discontinuity in the physical system upon perturbations), we can naively use the results from the representation theory of Poincar\'e group: for $s \geq 2$, massless fields have 2 physical DoFs, and massive fields have $2s+1$ physical DoFs. Then, the number of gauge redundancies (including traceless conditions) of a spin-$s$ massless field is
    \begin{equation}
        r_\text{massless}(4,s) = \binom{s+3}{s} - 2 = \frac{1}{6} (s+1) (s+2) (s+3)-2
    \end{equation}
    and that of a massive field is
    \begin{equation}
        r_\text{massive}(4,s) = \binom{s+3}{s} - (2s+1) = \frac{1}{6} s \left(s^2+6 s-1\right).
    \end{equation}
    To compare the number of redundancies with the number of conditions, we find the asymptotics of these at large $s$:
    \begin{equation}
        n(4,s) \sim \frac{1}{12}s^3, \qquad r_\text{massless}(4,s) \sim r_\text{massive}(4,s) \sim \frac{1}{6}s^3
    \end{equation}
    hence, we can see 
    \begin{equation}
        r_\text{massless}(4,s) > r_\text{massive}(4,s) > n(4,s)
    \end{equation}
    that the redundancies are enough to accommodate the focusing conditions. Moreover, in the large spin limit, the number of focusing conditions is only half of that of redundancies.
    
    In summary, we speculate that the focusing condition could be a \emph{necessary condition} for the physical consistency of higher-spin fields/theories but probably not a sufficient one.
    
    However, this is merely a speculation, and we have yet to investigate how consistent coupling between higher-spin fields and general curved backgrounds is achieved. Once these are clear, it would also be interesting to explore the meaning of these weight $w \geq 2$ components that are gauged away in general Killing horizon backgrounds.\footnote{E.g., how this matches to the flat-space picture that the temporal and longitudinal polarisations are gauged away.} Another shortfall is that we have only analysed the Type G conditions to draw an early guess. It is necessary to test the Type G condition against known examples of higher-spin Killing horizons. If it fails, then the Type S condition must be examined by explicitly calculating the form of $\mathcal{P}_2$ to determine whether our speculation is correct.
    
    \subsection{Outlook}
    At the current stage, the results of this paper are only abstract and exploratory: we have just followed our noses and investigated how the gravitational focusing on Killing horizons would go wrong when higher-spin fields are added to the Lagrangian. There are still many limitations and open problems that we hope to revisit at a later time:
    \begin{enumerate}
        \item  In order to solidify our speculation, a proper analysis of the relationship between the focusing condition and physical constraints on higher-spin fields/theories needs to be carried out. For example, it is interesting to examine how the indefinite terms are related to unphysical modes. So far, our discussions lack concrete examples. It would be desired to construct one and test out our theory. Also, it is important to study the structure of the indefinite term $\mathcal{P}_2$ in specific models. For instance, one could question our investigation on a de Sitter background because it has a cosmological horizon and is maximally symmetric. It could be more challenging to construct black hole examples. We hope to pursue these in future works.
        \item We assumed the theory involving higher-spin fields can be formulated with a Lagrangian description in order to use covariant phase space equations to study the structure of $E_{vv}$ and the generalised Raychaudhuri equation. However, some concrete models of massless higher-spin fields (e.g., Vasiliev higher-spin gravity) have no known Lagrangian description. We believe this is not a big problem because we can directly study the structure of equations of motion in any specific theory. However, there could be new difficulties in extracting the metric equations of motion from the higher-spin language.
        \item We have yet to explicitly include an infinite tower of higher-spin fields in our analysis. It would be crucial to examine the convergence of the summations involved in our calculations, especially in concrete models such as Vasiliev gravity. 
        \item We have invoked St\"uckelberg's trick to reduce the massive higher-spin fields into a tower of massless fields, and the null indices can essentially be addressed using the massless transversality condition. However, as said previously, the gauge parameters of these massless fields could appear in the entropy density extracted from the focusing equation, rendering the horizon entropy gauge-dependent. This could be related to the field redefinition (non)invariance of the entropy density-current. It also suggests that the indefinite terms might just be illusions: indefinite terms could result from a bad choice of field variable frames, and they could be eliminated by an appropriate field redefinition such as the St\"uckelberg process. Essentially, the gauge dependence and field redefinition non-invariance could be a manifestation that the higher-spin fields and their gauge symmetries spoil the conventional concept of geometry in terms of a metric. (See point 8 for a related discussion.)
        \item We required the higher-spin fields to be smooth on the entire horizon. In principle, one should be able to relax this assumption. E.g., a $p$-form field $A$ could be divergent on the bifurcation surface while keeping the physically observable curvature form $F = \dd{A}$ smooth. Such divergent gauge fields are crucial to allow for non-zero potentials on the horizon, enabling $\Phi \delta Q$-like terms in the first law of thermodynamics. Moreover, as a technical point, it will be a non-trivial generalisation for the boost weight analysis to include non-smooth field components.
        \item As presented in the paper, we used Gaussian null coordinates to carry out the analysis. It would be helpful to also understand the generalised Raychaudhuri equation in a fully covariant manner, e.g., using differential form languages or the Killing field analysis in \cite{Hollands:2024vbe}.
        \item For a thorough understanding of gravitational focusing, it will be crucial to extend the generalised Raychaudhuri equation off the horizon. In \cite{HVWYZ}, we will show that the generalised linear Raychaudhuri equation would hold away from the horizon at an affine distance of linear order in the perturbation parameter. To push these results further, one needs to show how the Raychaudhuri equation can be sensibly generalised non-linearly. Additional constraints would help, e.g., one can restrict the discussions to effective field theories (EFT) of gravity only. In EFTs, one has a non-perturbative second law \cite{Davies:2023qaa}, and it would be interesting to study how a similar procedure can be carried out to obtain a generalised non-linear Raychaudhuri equation and extend it away from the horizon. 
        \item We may also need to reassess our consideration of causal structures, as discussed in our previous work \cite{Wall:2024lbd}. When defining a null vector $k^a$, we assumed that the ``nullness'' is given by the metric only, i.e., $g_{ab}k^a k^b = 0$. However, in a general theory of gravity and matter, the causal structure should be given by the fastest propagation surfaces, which could differ from the metric light cones. We may need to study the characteristics equation of these surfaces to find the correct light-like direction to better discuss gravitational focusing and horizons. Another issue is that higher-spin gauge transformations could map metrics with different causal structures to one another (see \cite{Ammon:2011nk,Castro:2011fm} for the 3D cases), which could spoil the notion of causal horizons. It would be interesting to ask how universal the gravitational focusing is under these ambiguities in the metric. We will come back to this problem in the future.
        \item In $f(\text{Riemann})$ theory, the holographic entanglement entropy formula by Dong \cite{Dong:2013qoa} matches with the Wall entropy formula. It is conjectured that Dong and Wall entropies agree in any diffeomorphism-invariant theory. When higher-spin fields are present, the Wall entropy is ill-defined if the focusing condition is not satisfied. We can use this to test against the conjecture: if the Dong entropy can be defined regardless of the focusing condition (or its equivalent in the extremal surface picture), then we can immediately disprove the conjecture for theories with higher-spin fields in general. However, one could restrict the conjecture of equivalence to focusing-compliant theories only.\footnote{The author thanks Diandian Wang for discussions on this point.} (See \cite{Dong:2023bax} for an analysis of Dong entropy involving higher-spin fields, but it is from the perspective of renormalisation group flow.)
        \item Recently, in \cite{Wang:2024byb}, classical localised shockwave solutions have been found for higher-spin fields on black hole backgrounds. In holographic theories, these higher-spin shockwave solutions correspond to out-of-time-order correlators of the boundary CFT whose Lyapunov exponents exceed the chaos bound. Interestingly, the higher-spin shockwave solutions presented in \cite{Wang:2024byb} violate the Type G focusing condition on the horizon, at least in their chosen gauge. This could be evidence that the focusing condition is related to physical consistency conditions, such as unitarity. To further explore the relationship between the focusing condition and the chaos bound, one could check whether there is a focusing-compliant gauge for the shockwave solution. Or one could construct examples of shockwave solutions in some particular higher-spin theories and examine both the Type G and the Type S conditions. 
    \end{enumerate}

    \section*{Acknowledgements}

    The author is grateful to Alejandra~Castro, Zucheng~Gao, Gary~Horowitz, Veronika~Hubeny, Harvey~Reall, Jorge~Santos, Xi~Tong, Manus~Visser, Robert~Wald, Aron~Wall, Diandian~Wang, Zi-Yue~Wang, Wayne~Weng and Victor~Zhang for helpful discussions. The author especially thanks Aron~Wall for extensive and insightful conversations. The author is greatly indebted to Xiao~Chen, Jiali~Song and Unity~Leyou~Yan for their unconditional support during the completion of this work. The author is supported by an Internal Graduate Studentship of Trinity College, Cambridge. This work was supported in part by AFOSR grant FA9550-19-1-0260 ``Tensor Networks and Holographic Spacetime'' and grant NSF PHY-2309135 to the Kavli Institute for Theoretical Physics (KITP). The author is grateful for the hospitality of KITP and UCSB, where part of this work was completed.

    \appendix
    \section{An example of indefinite term}\label{app:toy-model}
    We illustrate the existence of the indefinite terms using a simple unphysical toy model. Consider a spin-2 field $\varphi_{ab}$ coupled to $R^{ab}$:
        \begin{equation}
            L = (16 \pi G)^{-1} R + \lambda \varphi_{ab} R^{ab}
        \end{equation}
        where $G$ is the Newton's constant. The off-shell EoMs are 
        \begin{equation}
            \mathcal{E}^{ab} = \lambda R^{ab}
        \end{equation}
        and
        \begin{equation}
            \begin{split}
                E_{ab} & = \frac{G_{ab}}{8 \pi G} + 2\chi_{(ab)} -  \lambda g_{ab} \varphi^{cd} R_{cd} + \lambda g_{ab} \nabla_{d}\nabla_{c}\varphi^{cd} -  \frac{1}{2} \lambda \nabla_{c}\nabla_{a}\varphi_{b}{}^{c} -  \frac{1}{2} \lambda \nabla_{c}\nabla_{a}\varphi^{c}{}_{b}\\
                & \quad -  \frac{1}{2} \lambda \nabla_{c}\nabla_{b}\varphi_{a}{}^{c} -  \frac{1}{2} \lambda \nabla_{c}\nabla_{b}\varphi^{c}{}_{a} + \frac{1}{2} \lambda \nabla_{c}\nabla^{c}\varphi_{ab} + \frac{1}{2} \lambda \nabla_{c}\nabla^{c}\varphi_{ba} 
            \end{split}
        \end{equation}
        where $G_{ab} = R_{ab} - \frac{1}{2} g_{ab} R$, and we have recovered $\chi_{ab} = \lambda R_{ac} \varphi_{b}{}^{c} + \lambda R_{ca} \varphi^{c}{}_{b}$. 

        At first order of perturbation, a GNC decomposition of the $vv$-component of the constraint form reads 
        \begin{equation}
            \begin{split}
                \delta (E_{vv} - \chi_{vv}) & \fheq - \partial_v \left(\frac{1}{\sqrt{h}} \partial_v \delta \left( \sqrt{h} \left( \frac{1}{8 \pi G} + 2 \lambda \varphi_{(uv)} \right) \right) + \lambda D_i \left(\delta \varphi\indices{_v^i} + \delta \varphi\indices{^i_v} + v \nabla^i \delta \varphi_{vv}\right)\right) \\
                & \qquad + \lambda \mathcal{L}_\xi \left(D_i \nabla^i \delta \varphi_{vv} - (\partial_v \bar K) \delta \varphi_{vv} \right) 
            \end{split}
        \end{equation}
        where $\bar K = (\sqrt{h})^{-1} \partial_u \sqrt{h}$ is the expansion in $u$-direction. Here, we clearly see the existence of an indefinite term: 
        \begin{equation}
            \mathcal{L}_\xi \mathcal{P}_2 = \frac{\lambda}{2 \pi} \mathcal{L}_\xi \left( (\partial_v \bar K) \delta \varphi_{vv} - D_i \nabla^i \delta \varphi_{vv} \right),
        \end{equation}
        which does not vanish in general since no physical constraints are imposed to set $\mathcal{L}_\xi \delta \varphi_{vv}$ to zero.

    \section{Focusing condition in 3D higher-spin black holes}\label{app:3d}
    Using an example, we demonstrate that in a well-defined higher-spin theory, there should be enough gauge symmetry to impose the correct gauge conditions that ensure focusing. 

    Consider a black hole in three-dimensional $\mathfrak{sl}(3,\mathbb{R}) \oplus \mathfrak{sl}(3,\mathbb{R})$ higher-spin gravity. Here, a spin-3 field $\varphi_{(abc)}$ is present. In a non-rotating stationary black hole example \cite{Gutperle:2011kf}, it is demonstrated that the following \emph{black hole gauge} can be chosen for the metric and the spin-3 field near the horizon $r=0$:
    \begin{align}
        \dd{s^2} & \approx - \kappa^2 r^2 \dd{t^2} + \dd{r^2} + g_{\phi \phi}(0) \dd{\phi^2}\\
        \varphi_{abc} \dd{x^a}\dd{x^b}\dd{x^c} & = \varphi_{\phi r r}(r) \dd{\phi} \dd{r^2} + \varphi_{\phi t t}(r) \dd{\phi} \dd{t^2} + \varphi_{\phi \phi \phi}(r) \dd{\phi^3}
    \end{align}
    where polar coordinates $(t,r,\phi)$ are used and $\kappa$ is the surface gravity which is constant. 

    Perform coordinate transformation to GNC $(u,v,\phi)$ using 
    \begin{equation}
        r = \sqrt{2 u v} \qquad \text{and} \qquad t = \frac{1}{2 \kappa} \log(v/u),
    \end{equation}
    we get 
    \begin{equation}
        g_{vv} = 0, \qquad \varphi_{vvv} = 0, \qquad \partial_u \varphi_{vvv} = 0
    \end{equation}
    near the horizon, and 
    \begin{equation}
        \varphi_{\phi v v} = \left(\pdv{r}{v}\right)^2 \varphi_{\phi rr} + \left(\pdv{t}{v}\right)^2 \varphi_{\phi t t} \fheq 0
    \end{equation}
    on the horizon, because the horizon is now labelled by $u = 0$ so $\pdv*{r}{v} \propto \sqrt{u} \fheq 0$, and $\varphi_{\phi t t} \fheq 0$ is required by the regularity of the horizon. 

    This example suggests that the $\mathfrak{sl}(3,\mathbb{R}) \oplus \mathfrak{sl}(3,\mathbb{R})$ higher-spin theory has enough gauge symmetry to satisfy the Type G focusing condition. Though the discussion is based on a stationary black hole solution, for consistency, the gauge symmetry should be respected even if non-stationary perturbations are switched on. So, the gauge condition can be protected, and we argue that the focusing condition continues to hold. Therefore,
    \begin{equation}
        \mathcal{L}_\xi \mathcal P_2 = 0,
    \end{equation}
    and the focusing theorem holds. The Wall entropy of the black hole can also be extracted.

    \newpage

    \bibliography{references}

\end{document}